\documentclass{article}
\usepackage{amsmath}
\usepackage{amssymb}
\usepackage{amsfonts}
\usepackage{bm}
\usepackage[dvips,xdvi]{color}
\usepackage[dvips,xdvi]{graphicx}
\usepackage{mciteplus}
\usepackage[compress,numbers]{natbib}

\begin{document}


\title{\bf Coarse-Grained Simulation Model for Crystalline Polymer Solids
by using Breakable Bonds}

\author{Takashi Uneyama{\textsuperscript{*}}\\
\\
  JST-PRESTO, and Department of Materials Physics, \\
Graduate School of Engineering,
  Nagoya University, \\
Furo-cho, Chikusa, Nagoya 464-8603,
  Japan \\
e-mail: uneyama@mp.pse.nagoya-u.ac.jp}
\date{}

\maketitle




\begin{abstract}
We propose a highly coarse-grained simulation model for crystalline polymer solids
with crystalline lamellar structures.
The mechanical properties of a crystalline polymer solid are mainly
determined by the crystalline lamellar structures. This means that
coarse-grained models rather than fine-scale molecular models are suitable to study
mechanical properties.
We model a crystalline polymer solid by using highly coarse-grained
particles, of which size is comparable to the crystalline layer thickness.
One coarse-grained particle consists of multiple subchains, and
is much larger than monomers.
Coarse-grained particles are connected by bonds to form a network structure.
Particles are connected by soft but ductile bonds, to form a rubber-like network.
Particles in the crystalline region are connected by hard but brittle
bonds.
Brittle bonds are broken when large deformations are applied.
We perform uniaxial elongation simulations based on our coarse-grained model.
As the applied strain increases, crystalline layers are broken into
pieces and non-affine and collective motions of broken pieces are observed.
Our model can successfully reproduce yield behaviors which are similar
to typical crystalline polymer solids.
\end{abstract}

%

\section{Introduction}
\label{introduction}

Crystalline polymers form hierarchical structures when cooled and solidified
from melt states\cite{Strobl-book,Young-Lovell-book}.
At the microscopic scale, crystalline lattice structures which depend
on monomer structures are formed.
At the mesoscopic scale, crystals form
layer-like structures. 
Crystalline and amorphous layers are stacked and form
crystalline lamellar structures.
At the larger scale, crystalline lamellae form spherulite structures.
Mechanical properties of crystalline polymer solids reflect such
hierarchical structures\cite{Lin-Argon-1994}.
A typical nominal stress-strain curve for uniaxial elongation of
high density polyethelene (HDPE), which is one of widely used
crystalline polymers, is shown in
Figure~\ref{stress_strain_hdpe}.
(See Appendix~\ref{experimental} for details of the experimental setting.)
If the strain is sufficiently small, a crystalline polymer solid
behaves as an elastic body. As the strain increases, the
stress-strain curve deviates from that for a linear elastic body,
and exhibits a local maximum which corresponds to the yield point.
After the yield point, stress decreases and inhomogeneous neck
formation and propagation are observed.
The stress is almost constant during this neck propagation process.
After the neck propagation is completed, the stress linearly increases
with strain (the strain hardening), and finally the sample breaks.
Although the details of stress-strain curves depend on various factors such
as monomer structures and crystallinity, the behaviors explained above
are qualitatively common for most of crystalline polymer solids\cite{Strobl-book,Young-Lovell-book}.

We expect that macroscopically observed mechanical behaviors
originate from microscopic and mesoscopic structural deformations.
Some theoretical models have been proposed to explain macroscopic
mechanical behaviors starting from microscopic or mesoscopic molecular
models\cite{Lin-Argon-1994,Bowden-Young-1974,Peterlin-1987,Nitta-Takayanagi-2000,Nitta-Takayanagi-2003}.
Different theoretical models based on different pictures give
similar macroscopic mechanical behaviors.
To judge which model is reasonable, simultaneous measurements for
structural and mechanical properties are required.
To experimentally study structural deformation behaviors,
scattering methods\cite{Tashiro-1993,Tashiro-Sasaki-2003,LopezBarron-Zeng-Schaefer-Eberle-Lodge-Bates-2017,Butler-Donald-1998,Jiang-Tang-Men-Enderle-Lilge-Roth-Gehrke-Rieger-2007,Millot-Seguela-Lame-Fillot-Rochas-Sotta-2017,Kishimoto-Mita-Ogawa-Takenaka-2020}
as well as spectroscopic methods\cite{Tashiro-1993,Tashiro-Sasaki-2003,LopezBarron-Zeng-Schaefer-Eberle-Lodge-Bates-2017,Siesler-1980,Song-Nitta-Nemoto-2003,Kida-Oku-Hiejima-Nitta-2015,Kida-Hiejima-Nitta-2021,Kida-2022}
are useful. By combining scattering and/or spectroscopic
measurements together with mechanical measurements,
we can study the relation between structural and mechanical properties.
From scattering experiments, we have structural information such as
chain conformations and orientations of crystals and lamellar structures. From spectroscopy, we have 
other structural informations such as orientations of chains and
stresses applied to polymer chains.
These informations are useful to discuss the structural deformations
under large deformation.
But even if we combine several experimental techniques, it is difficult
to precisely determine how hierarchical structures are deformed in a real space.

Simulations are also useful to study how structures are deformed
and how structural deformations are related to mechanical behaviors.
To study structural and mechanical behaviors by simulations,
molecular models have been utilized\cite{Lee-Rutledge-2011,Kim-Locker-Rutledge-2014,JabbariFarouji-Rottler-Lame-Makke-Perez-Barrat-2015,Yeh-Lenhart-Rutledge-Andzelm-2017,Higuchi-Kubo-2017,Olsson-intVeld-Andreasson-Bergvall-Jutemar-Petersson-Rutledge-Kroon-2018,Ranganathan-Kumar-Brayton-Kroger-Rutledge-2020,Hagita-Yamamoto-Saito-Abe-2024}.
We can directly observe real space structures in molecular simulations.
Molecular dynamics (MD) models are widely used because both crystalline
and amorphous structures can be naturally handled. In addition, nonequilibrium dynamics can be
simulated by imposing deformations to simulation boxes.
Various MD models, such as
all-atom models\cite{Olsson-intVeld-Andreasson-Bergvall-Jutemar-Petersson-Rutledge-Kroon-2018},
slightly coarse-grained united-atom models\cite{Kim-Locker-Rutledge-2014,Yeh-Lenhart-Rutledge-Andzelm-2017,Higuchi-Kubo-2017,Ranganathan-Kumar-Brayton-Kroger-Rutledge-2020,Hagita-Yamamoto-Saito-Abe-2024},
and more coarse-grained models\cite{JabbariFarouji-Rottler-Lame-Makke-Perez-Barrat-2015},
can be used to study crystalline polymer solids.
Although it is reported that MD simulations can reproduce yield behaviors,
there is serious limitation.
In typical MD simulations, length and time scales are small and
imposed strain rates become too high.
Typically just several layers in crystalline lamellar structures
are used in MD simulations. This means that the structural deformations
and non-affine motions at larger scales cannot be reproduced.
Typical strain rates in MD simulations are $\dot{\varepsilon} \approx (10^{7} \sim 10^{9}) \text{s}^{-1}$, which
is too high compared with typical values in experiments, $\dot{\varepsilon} \approx 10^{-2} \text{s}^{-1}$.
This discrepancy becomes serious when we want to compare MD simulation data with
experimental data, because both mechanical and structural behaviors
of crystalline polymers strongly depend on strain rate.
Simulation models which enable deformations with
much lower strain rates are demanding.

\begin{figure}[tb!]
 \centering
{\includegraphics[width=0.5\linewidth,clip]{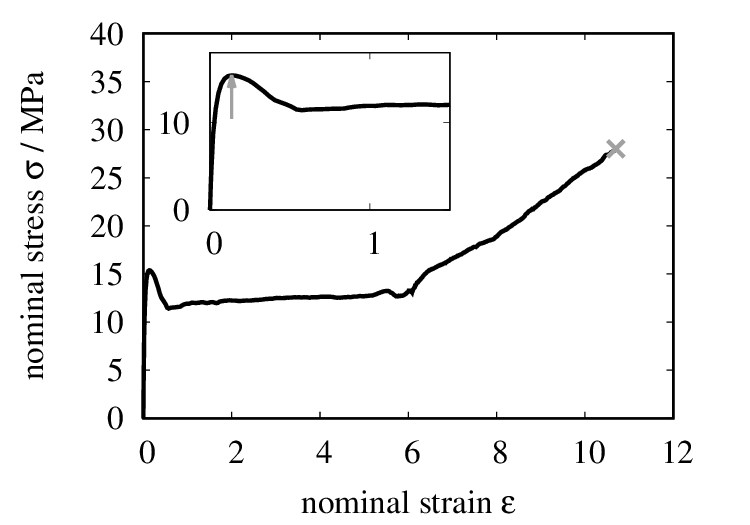}}
 \caption{A nominal stress-strain for high-density polyethylene (HDPE).
 Inset shows the data around the yield point.
 The gray arrow and cross symbol indicate the yield and break points, respectively.
}
 \label{stress_strain_hdpe}
\end{figure}

In this work, we propose a highly coarse-grained model for
crystalline polymer solids with crystalline lamellar structures.
We use a highly coarse-grained particles with a characteristic
size comparable to crystal and amorphous layers. These particles are much
larger than united-atoms, and thus the characteristic length and time scales
of our model are much larger than those of MD models. Our model
enables long-time simulations for large systems
with small numerical calculation costs.
To reproduce the yield behaviors, we employ 
the transient potential model\cite{Uneyama-2020,Uneyama-2022}.
In the transient potential model, potentials for coarse-grained particles are generally time-dependent
and fluctuate. In the case of crystalline polymers, bond potentials
in crystalline regions can be interpreted as the transient potential.
Bonds can be broken spontaneously if deformed largely.
We construct the potential model and dynamics model for the
highly coarse-grained particles. Then, we perform uniaxial elongation simulations
by using the constructed highly coarse-grained model.
Simulation results show that our model can reasonably reproduce both structural
and mechanical behaviors.
The crystalline layers are broken into pieces under large deformation.
The broken pieces move non-affinely and collectively.
The yield behavior associated with the breakage of crystalline layers
are observed.
Both structural and mechanical behaviors depend on the strain rate,
and the yield stress obeys the Eyring type relation.

\section{Model}
\label{model}
As explained in Sec.~\ref{introduction}, 
fine-scale MD simulations are not suitable when we are
interested in the long-time scale behaviors of crystalline polymer solids
with crystalline lamellar structures. Although fine-scale MD simulations
give informations on small-scale structures and dynamics, 
large-scale structures and dynamics are essential in many cases.
To overcome this problem, we employ highly coarse-grained model
where characteristic time and length scales are much larger than
those of MD models.
For mechanical behaviors, crystalline
lamellae structures (of the order of $10 \text{nm}$) are important rather than crystalline lattice
structures (of the order of $0.1 \text{nm}$).
Nitta and Takayanagi\cite{Nitta-Takayanagi-2000,Nitta-Takayanagi-2003} proposed that crystalline layers are broken
into small units called lamellar cluster units under large deformation.
Crystalline layers are broken into small units when they are deformed.
But different crystalline layers are connected by tie chains, so
multiple units form a collectively moving unit.
This scenario reasonably explains
yield and plastic flow behaviors of crystalline polymers.
Also, it is not sensitive to detailed monomer structures and
crystalline lattice structures.
Thus we expect that deformation and mechanical properties of 
crystalline polymers will be reproduced with a highly coarse-grained
model, of which characteristic scale is comparable to thicknesses of
crystalline and amorphous layers.

\begin{figure}[tb!]
 \centering
{\includegraphics[width=0.8\linewidth,clip]{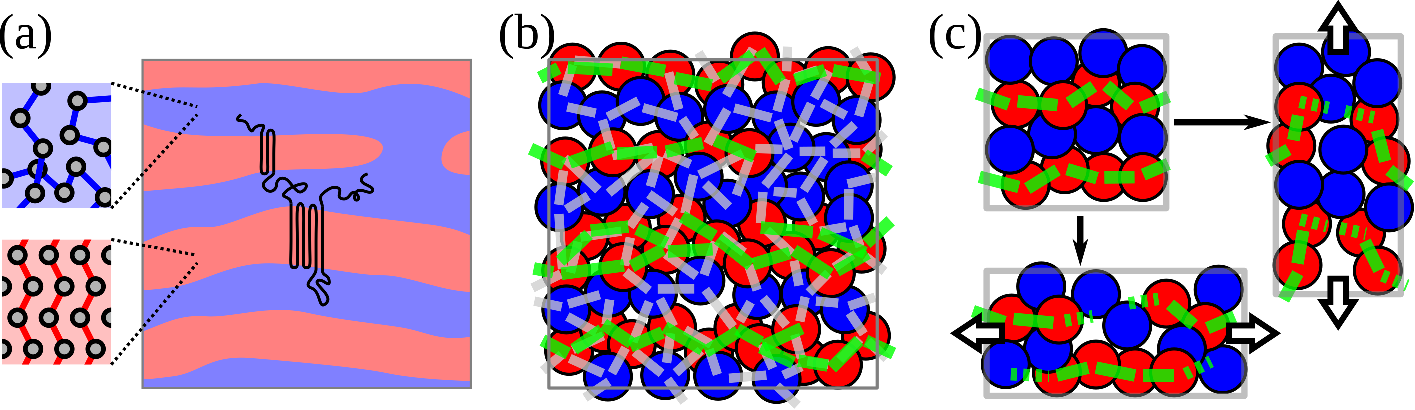}}
 \caption{Images of the coarse-grained model in this work.
 (a) A crystalline lamellar structure in a crystalline polymer solid.
 The lamellar structure consists of crystalline and amorphous
 regions (red and blue regions, respectively).
 The black curve shows a single polymer chain.
 At the microscopic scale,
 polymer chains in the crystalline region form a crystalline lattice,
 where as polymer chains in the amorphous region have no crystalline order.
 (b) Highly coarse-grained expression by using particles.
 Red and blue particles
 represent the crystalline and amorphous particles, respectively.
 Gray and green line segments represent bonds which connect particles.
 A green bond connect crystalline particles and can be broken under
 large deformation. A gray bond connect both crystalline and amorphous
 particles, and cannot be broken.
 (c) Coarse-grained structure under large deformation. When a large
 strain is applied, some bonds will be broken (dotted green line segments).
}
 \label{coarse_graining_image}
\end{figure}

We employ highly coarse-grained particles of which size is comparable to
the thickness of a crystalline layer in crystalline lamellae.
(These particles may be considered as the granular crystal layer
proposed as Strobl\cite{Strolb-2000}.)
We should construct interaction potentials for 
coarse-grained particles so that crystalline lamellar structures
are formed by the particles.
Intuitively, we can model one crystalline layer as two-dimensionally
packed and connected particles. One amorphous layer can be modeled
in a similar way. By stacking crystalline and amorphous layers,
we can form crystalline lamellae. But different crystalline layers
are connected by tie subchains\cite{Nitta-Takayanagi-2000,Nitta-Takayanagi-2003,Uneyama-Miyata-Nitta-2014}
and entanglements between loop subchains. Thus the stacked layers should be
connected by some bonds.
Thus the design of bond potentials will be important in our model.
The author derived a general coarse-grained
dynamics model in which potentials between coarse-grained particles
are time-dependent and fluctuate (the transient potential model)\cite{Uneyama-2020,Uneyama-2022}.
Based on the idea of the transient potential model, we consider
that the bond potential which connect crystalline particles can
change under large deformation.
Then some bonds will be spontaneously broken under large deformation,
and such breakage will lead macroscopic yield behaviors.
The transient potential can express such breakage in a simple and
physically natural way.
Figure~\ref{coarse_graining_image}
shows images of our coarse-grained model.
We do not consider the void formation, which is observed
in crystalline polymer solids. We assume that the system is incompressible
and assume that the Poisson's ratio is $\nu = 1/2$.
In what follows, we construct the highly coarse-gained model based on
this picture.

\subsection{Interaction Potential}
\label{interaction_potential}

We express the system as a network structure, in which particles are
connected by bonds. We introduce two particle types. One is for particles
in the crystalline region and another is for particles in the amorphous
region. We call these particles as crystalline and amorphous particles.
We model a particle as an elastic sphere. Particles are connected by bonds
to form a network structure. 
We design a simple interaction potential model for two types of
particles.

Before we consider crystalline layers, we virtually consider
a simple rubber-like network which consists only of amorphous particles.
This network should behave as an elastic solid even under large
deformation.
We expect a bond in this network can be stretched largely and
it cannot be broken.
It is soft but ductile.
We may call such a bond as a D-bond (ductile bond).

To model crystalline layers, we introduce a different type of bond.
Crystalline layers are hard but brittle. Thus we should consider
that breakage of bonds in the crystalline layers.
Namely, a bond in the crystalline region is not permanent but changes
according to its environment. Such a bond
can be interpreted as a nonequilibrium transient potential\cite{Uneyama-2020,Uneyama-2022}.
In the transient potential model, a transient potential for
coarse-grained particles relaxes if particle positions change.
We interpret this relaxation process as destruction of bonds,
and allow bonds can be broken during a dynamics simulation.
We may call such a bond as a B-bond (brittle or breakable bond).
It would be worth noting that the behavior of a B-bond
is qualitatively similar to the reactive force fields\cite{vanDuin-Dasgupta-Lorant-Goddard-2001}, except that
the characteristic length scale of the reactive force fields is
much smaller.
The particles in crystalline layers are connected both by D-bonds and
B-bonds.

We express the position of the $j$-th particle in the system as $\bm{r}_{j}$.
Each particle has a director which is expressed as a unit vector, and
we express the director of the $j$-th particle as $\bm{u}_{j}$.
As explained, particles are connected by B-bonds and D-bonds
to form a network structure.
We express the indices of two particles
connected by the $\alpha$-th D-bond as $S^{(\text{D})}_{\alpha,1}$ and $S^{(\text{D})}_{\alpha,2}$.
In a similar way, the particles with indices $S^{(\text{B})}_{\alpha,1}$ and $S^{(\text{B})}_{\alpha,2}$
are connected by the $\alpha$-th B-bond.
Now the state of the system can be expressed by the set of variables:
$\lbrace \bm{r}_{j} \rbrace$, $\lbrace \bm{u}_{j} \rbrace$, $\lbrace S^{(\text{D})}_{\alpha,p} \rbrace$,
and $\lbrace S^{(\text{B})}_{\alpha,p} \rbrace$.

All the particles interact each other via the short-range repulsive interaction potential.
We model a particle as an elastic sphere with a diameter $\sigma_{0}$. Then, the
interaction potential between particles can be modeled as the Hertzian
contact interaction potential\cite{Landau-Lifshitz-elasticity-book,Pamies-Cacciuto-Frenkel-2009}:
\begin{equation}
 \label{contact_potential}
 U_{\text{contact}}(\bm{r})  = 
  \begin{cases}
   \displaystyle  \epsilon \left( 1 - \frac{|\bm{r}|}{\sigma_{0}} \right)^{5/2} & (|\bm{r}| < \sigma_{0}), \\
   0 & (|\bm{r}| \ge \sigma_{0}) .
  \end{cases}
\end{equation}
Here, $\epsilon$ is a parameter which represents the interaction strength,
and $\bm{r}$ is the distance between particles.
Although $\epsilon$ can be related to the elastic modulus of particles,
in this work we just interpret $\epsilon$ as a phenomenological parameter.
Eq~\eqref{contact_potential} is purely repulsive, and we will have
a liquid of elastic spheres unless the repulsive strength is very strong
and the particle density is high\cite{Pamies-Cacciuto-Frenkel-2009}.
We expect that a D-bond will behave as an ideal chain\cite{Doi-Edwards-book}. Then the bond potential
for a D-bond becomes a simple harmonic bond potential:
\begin{equation}
 \label{dbond_potential}
 U_{\text{D-bond}}(\bm{r}) = \frac{1}{2} k_{\text{D-bond}} \bm{r}^{2} ,
\end{equation}
where $k_{\text{D-bond}}$ is the spring constant. 
We may relate
the spring constant to the average size $\bar{Q}$ of the corresponding
ideal chain as $k_{\text{D-bond}} = {3 k_{B} T} /{\bar{Q}^{2}} .$
Here, $k_{B}$ is the Boltzmann constant and $T$ is the temperature.
By combining potentials \eqref{contact_potential} and \eqref{dbond_potential},
we can construct an elastic solid which corresponds to a cross-liked
rubber.

We model the potentials for the B-bond so that
a crystalline layer behaves as an elastic plate under small deformation.
For this purpose, we combine three different potentials to express
one B-bond.
We model the B-bond stretch potential as the following shifted harmonic spring potential:
\begin{equation}
 \label{bbond_potential}
 U_{\text{B-bond}}(\bm{r};b) = \frac{1}{2} k_{\text{B-bond}} (|\bm{r}| - b)^{2} ,
\end{equation}
where $k_{\text{B-bond}}$ is a spring constant, and $b$ is the equilibrium
size of the B-bond. (The equilibrium size is not common for all the
B-bonds. The sizes are determined when B-bonds are formed.)
If particles are connected only by the B-bond stretch potential \eqref{bbond_potential},
we have no bending potential. To mimic the behavior of an elastic plate,
we should introduce additional potentials. 
Usually, such potentials are expressed as three body potential.
But the bending potential can be efficiently formulated 
as two body potential, by using directors in addition
to positions\cite{Noguchi-2011,Kuzkin-Asonov-2012,Kuzkin-Krivtsov-2017,Enomoto-Ishida-Doi-Uneyama-Masubuchi-2024}.
(This is why we introduced directors for coarse-grained particles.)
We introduce the
tilt and bending potentials:
\begin{align}
 \label{tilt_potential}
 U_{\text{tilt}}(\bm{r},\bm{u},\bm{u}')
 & = \frac{1}{2} k_{\text{tilt}} 
 \left[ \frac{(\bm{u} \cdot \bm{r})^{2}}{\bm{r}^{2}}
 +  \frac{(\bm{u}' \cdot \bm{r})^{2}}{\bm{r}^{2}}
 \right],  \\
 \label{bend_potential}
 U_{\text{bend}}(\bm{u},\bm{u}')
 & = \frac{1}{2} k_{\text{bend}} 
 \left[ 1 - (\bm{u} \cdot \bm{u}')^{2}
 \right] .
\end{align}
where $\bm{u}$ and $\bm{u}'$ are directors of two particles connected by
a B-bond, $k_{\text{tilt}}$ and $k_{\text{bend}}$ are strength
of tilt and bending potentials. Eqs~\eqref{tilt_potential} and \eqref{bend_potential} are symmetric under the
exchange of the sign of directors ($\bm{u} \to - \bm{u}$ and/or $\bm{u}' \to - \bm{u}'$).
The tilt potential \eqref{tilt_potential} drives directors to
be perpendicular to the bond, and the bending potential \eqref{bend_potential} drives directors
to be parallel to each other. Therefore, if we have a plate-like structure
of particles connected by B-bonds, the directors will have
the normal direction to the plate.

By combining eqs~\eqref{contact_potential}-\eqref{bend_potential}, the total potential energy of the system is given as
\begin{equation}
 \label{total_potential}
 \begin{split}
  & U(\lbrace \bm{r}_{j} \rbrace,\lbrace \bm{u}_{j} \rbrace, \lbrace S_{\alpha,p}^{(\text{D})} \rbrace, \lbrace S_{\alpha,p}^{(\text{B})} \rbrace ; \lbrace b_{\alpha} \rbrace) \\
  & = \sum_{j = 1}^{N} \sum_{k = j + 1}^{N} U_{\text{contact}}(\bm{r}_{j} - \bm{r}_{k})
  + \sum_{\alpha = 1}^{N_{\text{D}}} U_{\text{D-bond}}\left(\bm{r}_{S_{\alpha,1}^{(\text{D})}} - \bm{r}_{S_{\alpha,2}^{(\text{D})}}\right) \\
  & \qquad + \sum_{\alpha = 1}^{N_{\text{B}}} \Bigg[
  U_{\text{B-bond}}\left(\bm{r}_{S_{\alpha,1}^{(\text{B})}} - \bm{r}_{S_{\alpha,2}^{(\text{B})}}, b_{\alpha} \right) 
  + U_{\text{tilt}}\left(\bm{r}_{S_{\alpha,1}^{(\text{B})}} - \bm{r}_{S_{\alpha,2}^{(\text{B})}},\bm{u}_{S_{\alpha,1}^{(\text{B})}},\bm{u}_{S_{\alpha,2}^{(\text{B})}}\right) \\
  & \qquad + U_{\text{bend}}\left(\bm{u}_{S_{\alpha,1}^{(\text{B})}},\bm{u}_{S_{\alpha,2}^{(\text{B})}}\right) 
  \Bigg] .
 \end{split}
\end{equation}
Here, $b_{\alpha}$ represents the equilibrium bond size of the $\alpha$-th B-bond. As we explained, the B-bond size is not common. $b_{\alpha}$
is determined when the network is formed. To make the B-bonds consistent with locally disordered particle structures, $b_{\alpha}$ should depend on the
local structure. If a regular structure is used as the initial structure, a common value can be used for all the bonds.
(How $b_{\alpha}$ is determined
will be explained in Sec.~\ref{initial_structure}.)

\subsection{Dynamics Model}
\label{dynamics_model}

We assume that the inertia effect is sufficiently weak and negligible. Then
the dynamic equations for the particle positions and directors are given
as the overdampled Langevin equations. We employ the following Langevin
equations\cite{Tao-denOtter-Dhont-Briels-2006}:
\begin{equation}
 \label{langevin_equation_r}
 \frac{d\bm{r}_{j}(t)}{dt}
  = \bm{v}_{\text{ext}}(\bm{r}_{j}) - \frac{1}{\zeta} \frac{\partial U(\lbrace \bm{r}_{j} \rbrace,\lbrace \bm{u}_{j} \rbrace, \lbrace S_{\alpha,p}^{(\text{D})} \rbrace, \lbrace S_{\alpha,p}^{(\text{B})} \rbrace,\lbrace b_{\alpha} \rbrace) }{\partial \bm{r}_{j}}
  + \sqrt{\frac{2 k_{B} T}{\zeta}} \bm{w}(t) ,
\end{equation}
\begin{equation}
 \label{langevin_equation_u}
 \begin{split}
  \frac{d\bm{u}_{j}(t)}{dt}
  & = \bm{\omega}_{\text{ext}}(\bm{u}_{j}) - \frac{1}{\zeta'} \bm{P}(\bm{u}_{j}) 
  \cdot \frac{\partial U(\lbrace \bm{r}_{j} \rbrace,\lbrace \bm{u}_{j} \rbrace, \lbrace S_{\alpha,p}^{(\text{D})} \rbrace, \lbrace S_{\alpha,p}^{(\text{B})} \rbrace,\lbrace b_{\alpha} \rbrace) }{\partial \bm{u}_{j}} \\
  & \qquad + \frac{k_{B} T}{\zeta'} \frac{\partial}{\partial \bm{u}_{j}} \cdot \bm{P}(\bm{u}_{j})
  + \sqrt{\frac{2 k_{B} T}{\zeta'}} \bm{P}(\bm{u}_{j}) \cdot \bm{w}'(t) .
 \end{split}
\end{equation}
Here, $\bm{v}_{\text{ext}}(\bm{r})$ is the externally imposed velocity and rotational velocity,
$\zeta$ and $\zeta'$ are the translational and rotational friction coefficients,
$\bm{P}(\bm{u})$ is the projection matrix, and $\bm{w}(t)$ and $\bm{w}'(t)$ are
Gaussian white noises.
If the friction is modeled as the Stokes type drag for a spherical particle with the diameter $\sigma_{0}$,
the ratio of the rotational and translational friction coefficients are given as $\zeta' / \zeta = \sigma_{0}^{2} / 3$\cite{Landau-Lifshitz-fluid-book}.
The noises $\bm{w}(t)$ and $\bm{w}'(t)$ satisfy the following fluctuation-dissipation relation:
\begin{align}
 & \langle \bm{w}(t) \rangle = 0, \qquad
 \langle \bm{w}(t) \bm{w}(t') \rangle = \bm{1} \delta(t - t'), \\
 & \langle \bm{w}'(t) \rangle = 0, \qquad
 \langle \bm{w}'(t) \bm{w}'(t') \rangle = \bm{1} \delta(t - t'), \\
 & \langle \bm{w}(t) \bm{w}'(t) \rangle = 0.
\end{align}
Here, $\langle \dots \rangle$ represents the statistical average and $\bm{1}$ is the unit tensor.
The explicit expressions for forces and torques by individual interaction potentials are given in
Appendix~\ref{calculations_for_force_torque_and_stress}.

The projection matrix is defined as $\bm{P}(\bm{u}) = \bm{1} - \bm{u}\bm{u} / \bm{u}^{2}$.
This projection matrix extracts the part of the vector which is
perpendicular to the director $\bm{u}$, so that the director size will be
kept to be unity: $|\bm{u}_{j}(t)| = 1$. It satisfies $\bm{P}(\bm{u}) \cdot \bm{u} = 0$
and $\bm{P}(\bm{u}) \cdot \bm{P}(\bm{u}) = \bm{P}(\bm{u})$.
The third term in the right hand side of \eqref{langevin_equation_u}
is the spurious drift term which arises from the nature of the Ito type
stochastic differential equation\cite{Gardiner-book}.
This drift term becomes $(\partial / \partial \bm{u}) \cdot \bm{P}(\bm{u}) = - 2 \bm{u} / \bm{u}^{2}$,
and it works as the force which is parallel to the director.

As the external flow field, we employ the flow field by the homogeneous uniaxial deformation.
The velocity gradient for the uniaxial elongation is given as
\begin{equation}
 \label{velocity_gradient_tensor_uniaxial_elongation}
\bm{\kappa} =
\begin{bmatrix}
 - \dot{\varepsilon} / 2 & 0 & 0 \\
 0 & - \dot{\varepsilon} / 2 & 0 \\
 0 &0 & \dot{\varepsilon}
\end{bmatrix} ,
\end{equation}
where $\dot{\varepsilon}$ is the nominal strain rate. As explained,
the Poisson's ratio is
assumed to be $\nu = 1/2$ and thus the volume of the simulation box is kept unchanged.
From eq~\eqref{velocity_gradient_tensor_uniaxial_elongation}, we have the following external velocity and rotational velocity.
\begin{equation}
 \bm{v}_{\text{ext}}(\bm{r}) = \bm{\kappa} \cdot \bm{r}, \qquad
  \bm{\omega}_{\text{ext}}(\bm{u}) = \frac{1}{2} (\bm{\kappa} - \bm{\kappa}^{\mathrm{T}}) \cdot \bm{u} = 0.
\end{equation}
Here, the elongational flow does not directly affect the rotational motion
of a director, from a symmetry.

In addition to the dynamics described by the Langevin equations \eqref{langevin_equation_r} and \eqref{langevin_equation_u}, we incorporate
the effect of the bond breakage. A B-bond can be broken if it is largely
deformed. In this work, we simply employ the energy of the target B-bond
to judge the breakage. We assume that a B-bond can be broken immediately
if its energy
exceeds a critical value $E_{\text{crit}}$. The energy of the
$\alpha$-th B-bond is given as
\begin{equation}
 \label{bbond_energy}
 \begin{split}
   E_{\text{B},\alpha}
  & =
  U_{\text{B-bond}}\left(\bm{r}_{S_{\alpha,1}^{(\text{B})}} - \bm{r}_{S_{\alpha,2}^{(\text{B})}}; b_{\alpha}\right) 
  + U_{\text{tilt}}\left(\bm{r}_{S_{\alpha,1}^{(\text{B})}} - \bm{r}_{S_{\alpha,2}^{(\text{B})}},\bm{u}_{S_{\alpha,1}^{(\text{B})}},\bm{u}_{S_{\alpha,2}^{(\text{B})}}\right) \\
  & \qquad + U_{\text{bend}}\left(\bm{u}_{S_{\alpha,1}^{(\text{B})}},\bm{u}_{S_{\alpha,2}^{(\text{B})}}\right) .
 \end{split}
\end{equation}
If $E_{\text{B},\alpha} > E_{\text{crit}}$ at a certain time, then the $\alpha$-th B-bond is broken.
The pair $S_{\alpha,1}^{(\text{B})}$ and $S_{\alpha,2}^{(\text{B})}$ 
is simply removed from the list of indices for B-bonds.
In our model, we do not reconstruct the broken B-bonds.
There is no reverse process, and thus the dynamics is nonequilibrium.

\subsection{Stress}
\label{stress}

To study mechanical properties, we need to calculate the stress.
The stress of the system can be calculated by considering a small virtual deformation
(the virtual work method)\cite{Doi-Edwards-book}.
We virtually apply the following
small deformation to the system:
$\bm{r}_{j} \to \bm{r}_{j} + \bm{E} \cdot \bm{r}_{j}$ and $\bm{u}_{j} \to \bm{u}_{j} + (\bm{E} + \bm{E}^{\mathrm{T}}) \cdot \bm{u}_{j} / 2$,
where $\bm{E}$ is the deformation gradient tensor.
We assume that the deformation is sufficiently small and
B-bonds are not broken.

The total potential energy \eqref{total_potential} is changed by this virtual deformation.
If we express the stress tensor as $\bm{\sigma}$,
the energy of the system should be increased by
$V \bm{\sigma} : \bm{E}$ ($V$ is the volume of the system),
when the strain is small and terms upto the first order in $\bm{E}$
are considered.
Then the difference of the potential energies before and after the virtual deformation
can be related to be the stress tensor $\bm{\sigma}$:
\begin{equation}
 \begin{split}
  & U(\lbrace \bm{r}_{j} + \bm{E} \cdot \bm{r}_{j} \rbrace,\lbrace \bm{u}_{j} + (\bm{E} + \bm{E}^{\mathrm{T}}) \cdot \bm{u}_{j} / 2 \rbrace, \lbrace S_{\alpha,p}^{(\text{D})} \rbrace, \lbrace S_{\alpha,p}^{(\text{B})} \rbrace; \lbrace b_{\alpha} \rbrace) \\
  & - U(\lbrace \bm{r}_{j} \rbrace,\lbrace \bm{u}_{j} \rbrace, \lbrace S_{\alpha,p}^{(\text{D})} \rbrace, \lbrace S_{\alpha,p}^{(\text{B})} \rbrace; \lbrace b_{\alpha} \rbrace) 
 = V \bm{\sigma} : \bm{E}. 
 \end{split}
\end{equation}
From the changes of the two-body potential energies by eqs~\eqref{contact_potential}-\eqref{tilt_potential}, we have the following
stress contributions for the contact, D-bond, B-bond, and tilt potentials:
\begin{align}
 \label{contact_stress}
 V \bm{\sigma}_{\text{contact}}(\bm{r})
  & =   \begin{cases}
   \displaystyle  \frac{5\epsilon}{2 \sigma_{0}} \left( 1 - \frac{|\bm{r}|}{\sigma_{0}} \right)^{3/2} \frac{\bm{r}\bm{r}}{|\bm{r}|} & (|\bm{r}| < \sigma_{0}), \\
   0 & (|\bm{r}| \ge \sigma_{0}) ,
     \end{cases}, \\
 \label{dbond_stress}
  V \bm{\sigma}_{\text{D-bond}}(\bm{r}) 
 & = k_{\text{D-bond}} \bm{r} \bm{r} , \\
 \label{bbond_stress}
  V \bm{\sigma}_{\text{B-bond}}(\bm{r},b) 
 & = k_{\text{B-bond}} (|\bm{r}| - b) \frac{ \bm{r} \bm{r}}{|\bm{r}|} , \\
 \begin{split}
 \label{tilt_stress}
  V \bm{\sigma}_{\text{tilt}}(\bm{r},\bm{u},\bm{u}') 
  & =
  k_{\text{tilt}}  \bigg[  
  \left(\frac{\bm{u} \cdot \bm{r}}{\bm{r}^{2}}\right)
  \frac{\bm{u} \bm{r} + \bm{r} \bm{u}}{2} 
  + \left(\frac{ \bm{u}' \cdot \bm{r}}{\bm{r}^{2}}\right)
  \frac{\bm{u}' \bm{r} + \bm{r} \bm{u}'}{2} \\
  & \qquad - \left[  \frac{ (\bm{u} \cdot \bm{r})^{2}}{\bm{r}^{2}}
  + \frac{ (\bm{u}' \cdot \bm{r})^{2}}{\bm{r}^{2}}
  \right]   \frac{\bm{r} \bm{r}}{\bm{r}^{2}}
 \bigg].
 \end{split}
\end{align}
(See Appendix~\ref{calculations_for_force_torque_and_stress} for detailed
calculations.)
The bending potential \eqref{bend_potential} does not contribute to the stress.
The total stress tensor of the system $\bm{\sigma}$ is given as the sum of
contributions by eqs \eqref{contact_stress}-\eqref{tilt_stress}:
\begin{equation}
 \begin{split}
  & \bm{\sigma}(\lbrace \bm{r}_{j} \rbrace,\lbrace \bm{u}_{j} \rbrace, \lbrace S_{\alpha,p}^{(\text{D})} \rbrace, \lbrace S_{\alpha,p}^{(\text{B})} \rbrace ; \lbrace b_{\alpha} \rbrace) \\
  & = \frac{1}{V} \Bigg[ \sum_{j = 1}^{N} \sum_{k = j + 1}^{N} \bm{\sigma}_{\text{contact}}(\bm{r}_{j} - \bm{r}_{k}) 
  + \sum_{\alpha = 1}^{N_{\text{D}}} \bm{\sigma}_{\text{D-bond}}\left(\bm{r}_{S_{\alpha,1}^{(\text{D})}} - \bm{r}_{S_{\alpha,2}^{(\text{D})}}\right) \\
  & \qquad + \sum_{\alpha = 1}^{N_{\text{B}}} \left[
  \bm{\sigma}_{\text{B-bond}}\left(\bm{r}_{S_{\alpha,1}^{(\text{B})}} - \bm{r}_{S_{\alpha,2}^{(\text{B})}}; b_{\alpha}\right) 
  + \bm{\sigma}_{\text{tilt}}\left(\bm{r}_{S_{\alpha,1}^{(\text{B})}} - \bm{r}_{S_{\alpha,2}^{(\text{B})}},\bm{u}_{S_{\alpha,1}^{(\text{B})}},\bm{u}_{S_{\alpha,2}^{(\text{B})}}\right)  \Bigg] \right] .
 \end{split}
\end{equation}
In the case of the uniaxial elongation, $\sigma_{zz} - (\sigma_{xx} + \sigma_{yy}) / 2$
works as the elongational stress. The nominal stress is calculated as
\begin{equation}
 \sigma = \frac{\sigma_{zz} - (\sigma_{xx} + \sigma_{yy}) / 2}{1 + \varepsilon},
\end{equation}
where $\varepsilon$ is the nominal strain.

\section{Simulation}
\label{simulation}

\subsection{Parameters and Algorithms}
\label{parameters_and_algorithms}

We employ the dimensionless units by setting $\sigma_{0} = 1$, $\epsilon = 1$, and
$\tau = \zeta \sigma_{0}^{2} / \epsilon$ = 1. We express the
temperature in the dimensionless energy unit as $T_{\text{eff}} = k_{B} T$.
We use a cubic box with the periodic boundary condition. The initial
box length is $L = 32$ (the volume is $V = 32^{3}$)
and the total number of particles is $65536$ (the
particle density is $2$). The potential parameters are set as
$k_{\text{D-bond}} = 0.01, k_{\text{B-bond}} = 10,$ and
$k_{\text{tilt}} = k_{\text{bend}} = 1$.
The rotational friction coefficient and
the effective temperature are set as $\zeta' = 1$ and $T_{\text{eff}} = 0.01$.
The critical B-bond energy is set as $E_{\text{crit}} = 0.5$.
(The potential parameters are tuned so that the model mimics realistic
crystalline polymers. See Appendix \ref{estimates_for_potential_parameters}.)

To integrate the Langevin equations \eqref{langevin_equation_r}
and \eqref{langevin_equation_u}, we employ the second order stochastic
Runge-Kutta scheme\cite{Honeycutt-1992}.
To generate the random numbers, the Mersenne-Twister method
\cite{Matsumoto-Nishimura-1998} and the Box-Muller method
\cite{Devroye-book} are utilized.
The time step size is $\Delta t = 0.05$.
The nominal strain rate $\dot{\varepsilon}$ is changed from $0.0001$ to
$0.02$. The elongation simulations are performed upto $\varepsilon = 4$.
During the elongation simulations, energies of B-bonds are
monitored at each step and we break B-bonds with higher energies than $E_{\text{crit}}$.

\subsection{Initial Structure}
\label{initial_structure}

We need to prepare initial
crystalline lamellae structures. In reality, the crystalline
lamellae structures are spontaneously formed by cooling polymer melts.
Crystals nucleate and grow, and then they form stacked crystalline lamellae.
However, direct simulations for such a structural formation process
are hard, especially in a particle-based coarse-grained model like ours.
(Much more coarse-grained models such as the cellular automaton model\cite{Raabe-2004}
will be required to simulate large-scale crystallization processes.)
In this work, we prepare crystalline lamellar
structures by an unphysical yet simple method, instead of mimicking
realistic cooling processes.

The preparation of the initial structure is done by four
steps. The first step is to prepare a liquid of elastic particles.
We put the particles without any bonds randomly into the simulation box.
Then a dynamics simulation based on the Langevin equations
are performed for a short time, to equilibrate
particle positions. (The strain rate is set to $\dot{\varepsilon} = 0$
for equilibration processes.)
The second step is to form D-bonds. After
the equilibration at the first step, we form D-bonds. We scan
particle pairs of which distance is less than the cut-off
distance $r_{\text{D,cut}}$. If the distance between two candidate
particles is $r$, a D-bond is formed for the pair by the Fermi-Dirac type
probability
\begin{equation}
 \label{dbond_probability}
 P_{\text{D-bond}} = \frac{1}{1 + \exp(E_{\text{D}} - \mu_{\text{D}})},
\end{equation}
where $E_{\text{D}} = k_{\text{D-bond}} r^{2} / 2$ is the energy
of a D-bond and $\mu_{\text{D}}$ is the chemical potential.
If the energy is much lower than $\mu_{\text{D}}$, the D-bond formation
probability is almost $1$ whereas if it is much higher than $\mu_{\text{D}}$,
the probability is almost $0$. We set $r_{\text{D,cut}} = 2.5$ and $\exp(-\mu_{\text{D}}) = 0.1$.
After scanning all the pairs and form D-bonds, 
a short-time dynamics simulation is performed
again for equilibration. After this equilibration, we have an equilibrated elastic
network structure.

The third step is to set particle types.
We give a field which describes the crystallinity distribution, $\chi(\bm{r})$,
and set particle types according to this field. We interpret $\chi(\bm{r})$
as the probability that the particle type at position $\bm{r}$ is the
crystal type. We consider the case where the lamellae are stacked
parallel to the $z$-axis.
We use the following hypothetical field as the crystallinity field:
\begin{align}
 \label{crystallinity_field}
  \chi(\bm{r}) & = 
 \begin{cases}
  \displaystyle \frac{1}{2}
  \left[ 1 + \tanh\left(\frac{\tilde{z}(\bm{r})}{\xi} \right) \right] & (0 \le \tilde{z}(\bm{r}) < \bar{\chi} D / 2), \\
  \displaystyle \frac{1}{2}
  \left[ 1 + \tanh\left(\frac{\bar{\chi} D - \tilde{z}(\bm{r})}{\xi} \right) \right] & (\bar{\chi} D / 2 \le \tilde{z}(\bm{r}) < (1 + \bar{\chi}) D / 2), \\
  \displaystyle \frac{1}{2}
  \left[ 1 + \tanh\left(\frac{\tilde{z}(\bm{r}) - D}{\xi} \right) \right] & ((1 + \bar{\chi}) D / 2 \le \tilde{z}(\bm{r}) < 1),
\end{cases} \\
\label{crystallinity_field_distance}
 \tilde{z}(\bm{r}) & = r_{z} - D \lfloor r_{z} / D \rfloor,
\end{align}
where $r_{z}$ is the $z$-component of the position, $D$ is the 
long period (lamellar spacing), $\bar{\chi}$ is the average crystallinity, and $\xi$ is
the thickness of the crystal-amorphous interface. $\lfloor \dots \rfloor$
represents the floor function.
The director for crystalline particles is set to 
$\bm{u} = [0,  0,  1]$, which is the normal to the lamellae.
We set $D = 1.6$, $\bar{\chi} = 0.5$ and $\xi = 0.05$.
($20$ crystalline layers with sharp interfaces are formed in a simulation box.)
The D-bonds are not changed by this particle type setting.
Namely, a crystalline particle can be connected to surrounding 
crystalline and amorphous particles by D-bonds.
The fourth step is the formation of B-bonds for pairs of particles
in the crystalline region.
We scan crystal particles pairs which is within the cut-off distance
$r_{\text{B,cut}}$. In addition to this cut-off, we use another cut-off
for directors. If we describe the distance between the particles
as $\bm{r}$ and the director $\bm{u}$, $|\bm{r} \cdot \bm{u}| < r'_{\text{B,cut}}$
should be satisfied. For the particle pairs within the cut-off,
B-bonds are formed by the following probability:
\begin{equation}
 \label{bbond_probability}
 P_{\text{B-bond}}  = \frac{1}{1 + \exp(E_{\text{B}} - \mu_{\text{B}})}.
\end{equation}
Here, $E_{\text{B-bond}} = U_{\text{tilt}}(\bm{r},\bm{u},\bm{u})$ is the B-bond energy
and $\mu_{\text{B}}$ is the chemical potential.
When the $\alpha$-th B-bond is formed, $b_{\alpha}$ is set to be the
same as the distance; $b_{\alpha} = |\bm{r}|$.
We set $r_{c} = 1.5$, $r_{c}' = 0.5$, and $\exp(-\mu_{\text{B}}) = 100$.
($-\mu_{B}$ is set to be a large value, in order to enhance formation of
B-bonds.)
After B-bonds are formed, a short-time simulation is performed
to equilibrate the system. After this final equilibration, we have
the initial structure for the elongation simulations.
The number densities of B-bonds and D-bonds are about $2.7$ and $2.2$,
respectively. (A particle connected to zero or one
bond cannot carry the stress of the network.
The fraction of particles which is not connected to any bond
or to a single bond is about $0.05$.
The fraction of particles which is not connected to
the D-bond or connected only to a single D-bond is about $0.19$.)

In the initial structure formed by the procedure shown above,
a pair of particles can be connected both by a D-bond and a B-bond.
In such a case, both the forces by the D-bond and the B-bond are applied
to particles. Because the force by the B-bond is much larger than
that by the D-bond, the contribution of the D-bond will be almost
negligible unless the B-bond is broken.

The deformation behaviors and mechanical properties depend on the
elongation direction relative to the lamellar stacking direction.
We change the lamellar stacking direction to perform simulations with
different elongation directions. The angle between $z$-axis and
the lamellar stacking direction $\theta$ is varied as $\theta = 0^{\circ},
15^{\circ},30^{\circ},45^{\circ},60^{\circ},75^{\circ},$ and $90^{\circ}$.
Here, it would be fair to mention that the periodic lamellar structure
is not compatible with the periodic boundary condition in some cases.
Thus we have some mismatches
of lamellar structures around box boundaries. Fortunately, our simulation
box is sufficiently large, and effects of such mismatches are practically negligible.

\section{Results}
\label{results}

\subsection{Structural Changes During Elongation}
\label{structural_changes_during_elongation}

\begin{figure}[tb!]
 \centering
{\includegraphics[width=\linewidth,clip]{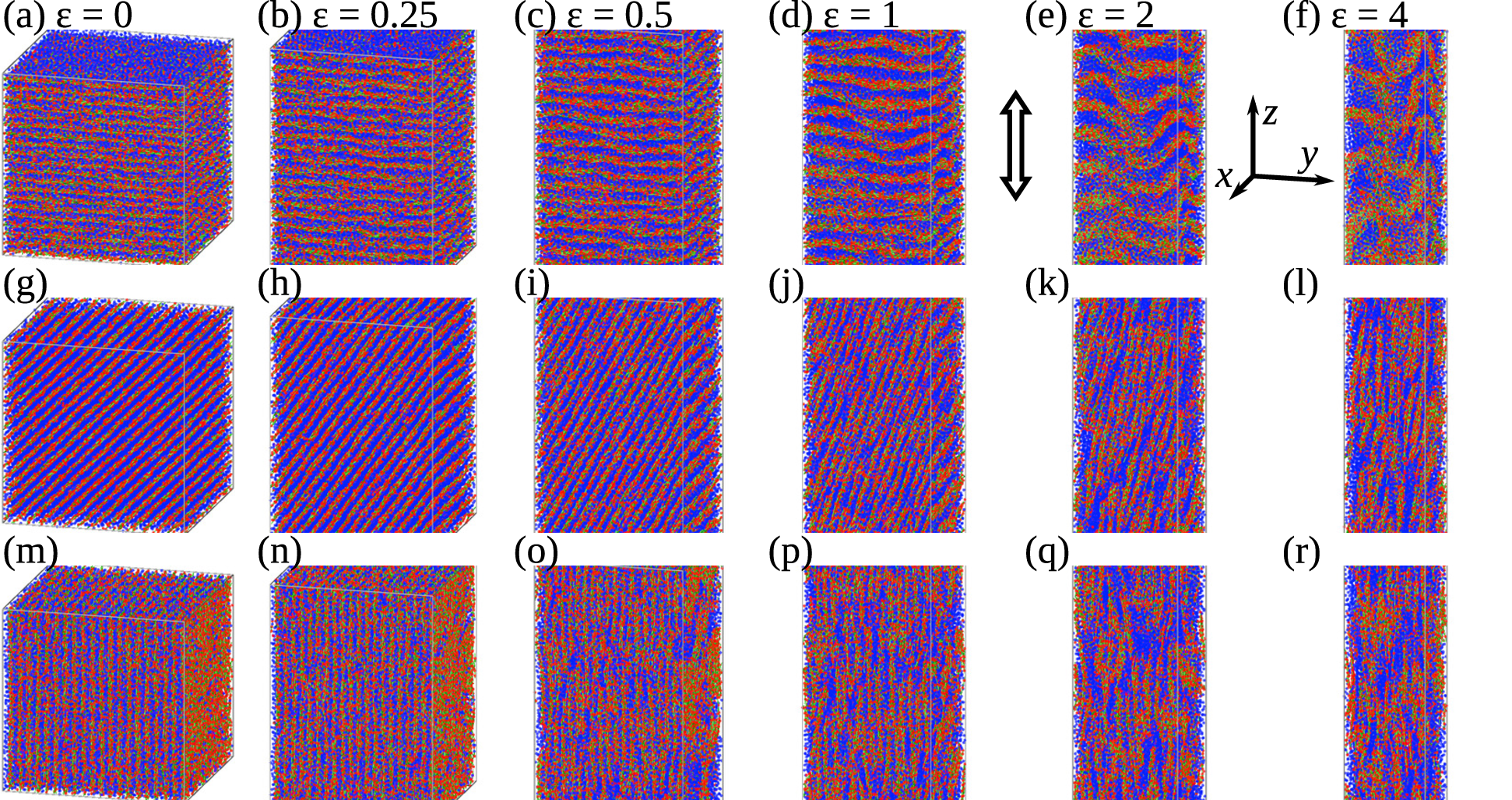}}
 \caption{Snapshots of elongation simulations of crystalline polymer
 solid by the coarse-grained model. Red and blue circles represent
 coarse-grained crystalline and amorphous particles, respectively.
 (The particle type is not changed even if all the B-bonds attached
 to a particle are broken.)
 Green line segments represent B-bonds which can be broken when largely
 deformed. The white double arrow shows the elongation direction ($z$-direction).
 The lamellar stacking directions are 
 (a)-(f) $\theta = 0^{\circ}$, 
 (g)-(l) $\theta = 45^{\circ}$, and
 (m)-(r) $\theta = 90^{\circ}$.
 Snapshots at different nominal strains ($\varepsilon = 0, 0.25, 0.5, 1, 2,$ and $4$) are shown.
 The nominal strain rate is $\dot{\varepsilon} = 0.002$.
}
 \label{snapshots_different_directions}
\end{figure}

Figure~\ref{snapshots_different_directions} shows typical snapshots
of elongation simulations based on our coarse-grained model.
(See also Supporting Information.)
If the strain is small $(\varepsilon = 0.25)$ (Figure~\ref{snapshots_different_directions}(b), (h), and (n)), 
we find that the lamellar
structures are almost the same as the initial structures. This means
that the deformation is almost affine. But as the strain increase,
we observe non-affine deformations. In the case of $\theta = 0^{\circ}$,
crystalline layers start to undulate around $\varepsilon = 0.5$
(Figure~\ref{snapshots_different_directions}(c)).
This can be interpreted as the
buckling of elastic plates\cite{Landau-Lifshitz-elasticity-book,Makke-Perez-Lame-Barrat-2012}.
As the strain is further increased, the undulation grows and finally we
observe that crystalline layers form plastic hinges
(Figure~\ref{snapshots_different_directions}(e) and (f)).
The motion of crystalline layers seem to be somewhat correlated.
In the case of $\theta = 90^{\circ}$, stretched crystalline layers
start to be tore and break into pieces (Figure~\ref{snapshots_different_directions}(o)).
The size of pieces of crystalline
layers are not largely changed even if the strain increases further.
Instead, pieces collectively rearrange their positions
(Figure~\ref{snapshots_different_directions}(p)-(r)).
In the case of $\theta = 45^{\circ}$, crystalline layers are rotated and
becomes parallel to the elongation direction. Then layers are broken
into pieces in a similar way to the case of $\theta = 90^{\circ}$.
In all the cases, the crystalline layers are broken and the structural
deformations are non-affine. As we show later, the breakage of the
crystalline layers can be observed as yield behaviors from the mechanical view point. (Due to the highly coarse-grained nature of the model,
we cannot study the microscopic mechanisms such as the crystalline slip
by our model.)

\begin{figure}[tb!]
 \centering
{\includegraphics[width=\linewidth,clip]{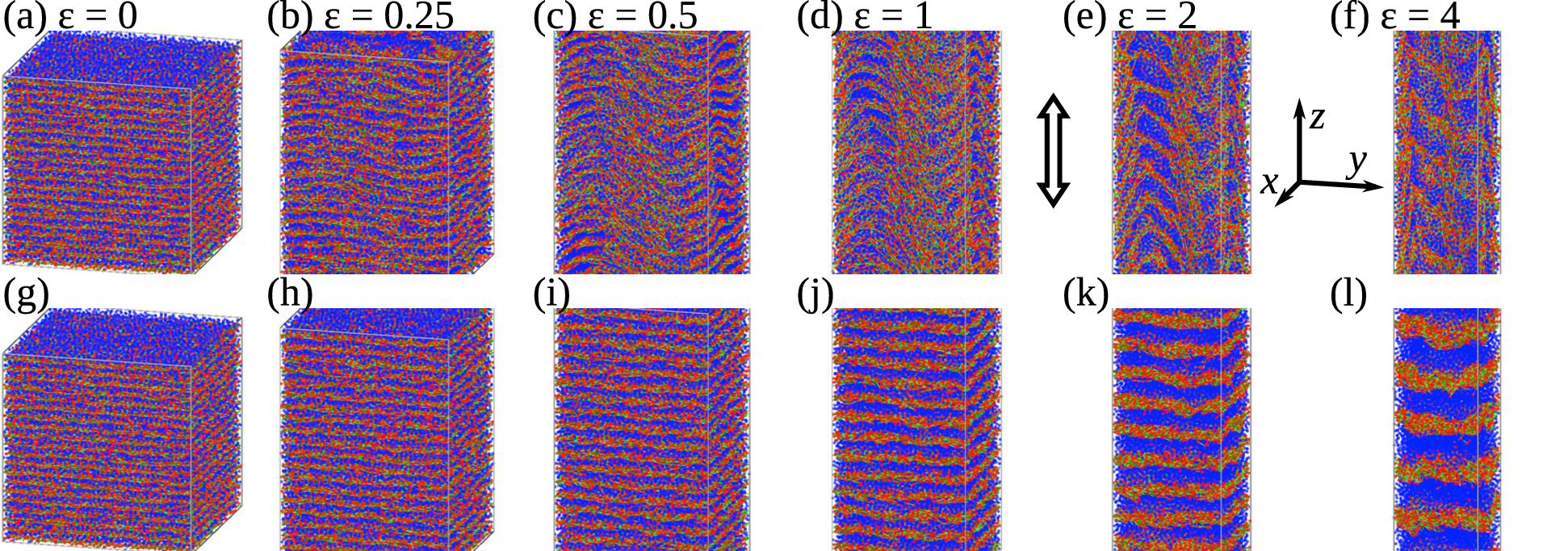}}
 \caption{Snapshots of elongation simulations with different nominal strain rates.
 The lamellar stacking direction is $\theta = 0^{\circ}$.
 (The visualization is the same as explained in Figure~\ref{snapshots_different_directions}.)
 (a)-(f) $\dot{\varepsilon} = 0.0002$, and
 (g)-(l) $\dot{\varepsilon} = 0.02$.
 (Figure~\ref{snapshots_different_directions}(a)-(f) corresponds to
 $\dot{\varepsilon} = 0.002$ with the same direction.)
}
 \label{snapshots_rate_dependence_0}
\end{figure}

\begin{figure}[tb!]
 \centering
{\includegraphics[width=\linewidth,clip]{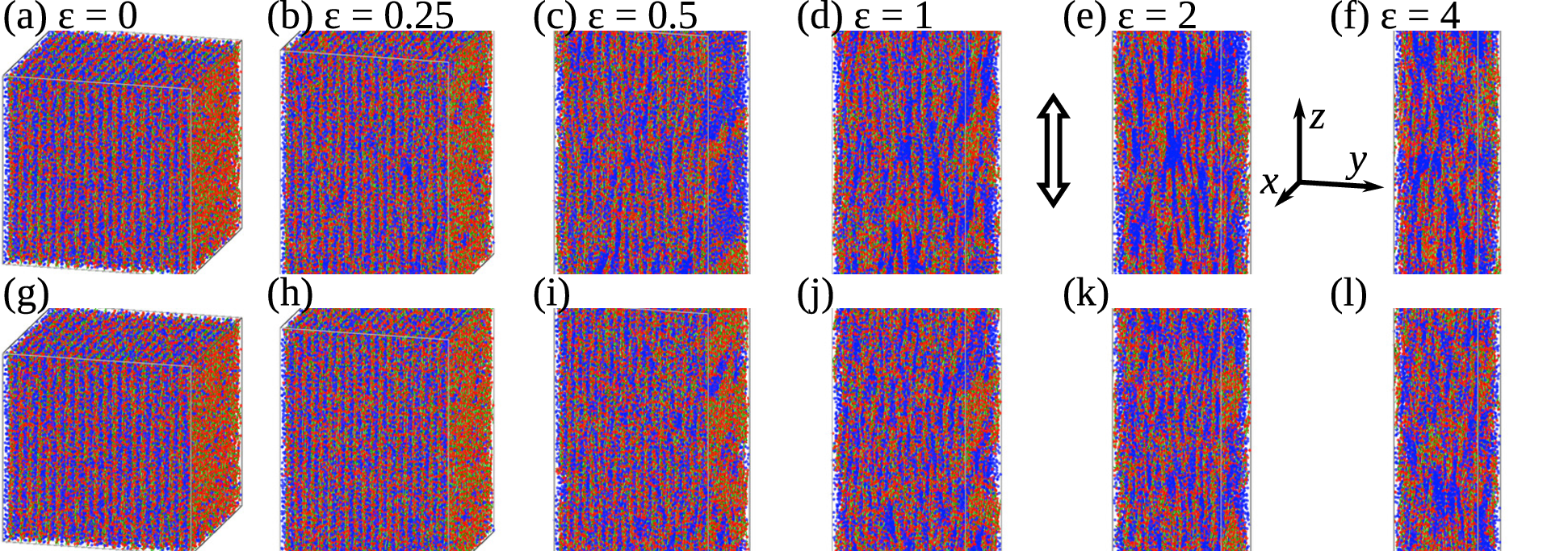}}
 \caption{Snapshots of elongation simulations with different nominal strain rates.
 The lamellar stacking direction is $\theta = 90^{\circ}$.
 (The visualization is the same as explained in Figure~\ref{snapshots_different_directions}.)
 (a)-(f) $\dot{\varepsilon} = 0.0002$, and
 (g)-(l) $\dot{\varepsilon} = 0.02$.
 (Figure~\ref{snapshots_different_directions}(m)-(r) corresponds to
 $\dot{\varepsilon} = 0.002$ with the same direction.)
}
 \label{snapshots_rate_dependence_90}
\end{figure}

Because the characteristic length scale of structures observed in
Figure~\ref{snapshots_different_directions} is larger than
the particle size (which is the unit length scale of the model), we expect that
the characteristic time scale of these structures become much larger
than the unit time scale. Thus the structural change is expected
to depend on the strain rate $\dot{\varepsilon}$, even if the
lamellar stacking direction is the same.
Figures~\ref{snapshots_rate_dependence_0} and
\ref{snapshots_rate_dependence_90} show the snapshots for $\theta = 0^{\circ}$
and $90^{\circ}$ with two different strain rates $\dot{\varepsilon} = 0.0002$ and $0.02$.
(The data for $\dot{\varepsilon} = 0.002$ are found in Figure~\ref{snapshots_different_directions}.
See also Supporting Information.)
We observe that the structural change strongly depends on the strain
rate, as expected. For the small strain rate ($\dot{\varepsilon} = 0.0002$), 
the non-affine deformation seems to be enhanced and we observe
very large-scale structures.
On the other hand, for the large strain rate ($\dot{\varepsilon} = 0.02$),
the large-scale structural change is rather suppressed. The non-affine
deformation is observed but the characteristic length scale is not large.

\begin{figure}[tb!]
 \centering
{\includegraphics[width=0.5\linewidth,clip]{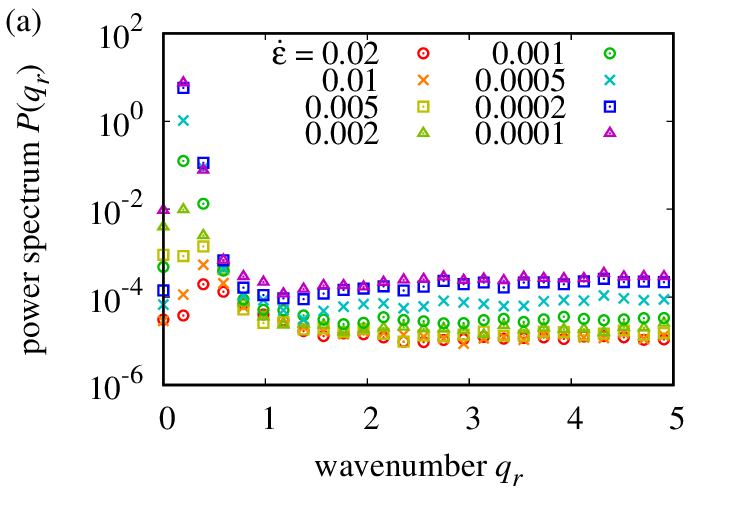}%
\includegraphics[width=0.5\linewidth,clip]{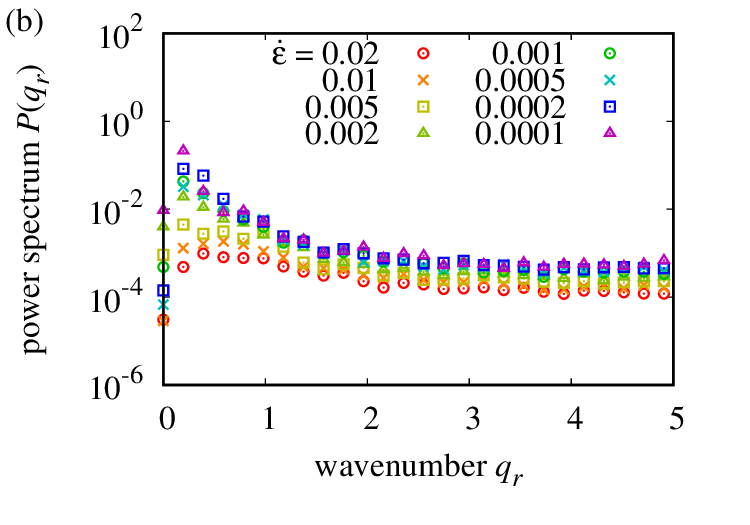}}
 \caption{Power spectra data for the non-affine displacement field in the $z$-direction
 on the $xy$-plane, $P(q_{r})$. (a) $\theta = 0^{\circ}$ and (b) $\theta = 90^{\circ}$.
}
 \label{nonaffine_deformation_spectra_0_90}
\end{figure}

To study the non-affine deformation in detail, here we analyze the
non-affine displacement field data.
We define the non-affine displacement field in the $z$-direction as
\begin{equation}
 \label{nonaffine_displacement_field_z}
 \Delta_{z}(\bm{r},t) = \sum_{j = 1}^{N} \delta(\bm{r} - \bm{r}_{j})
  [r_{j,z}(t) - (1 + \dot{\varepsilon} t) r_{j,z}(0)],
\end{equation}
where $r_{j,z}(t)$ is the $z$-component of the position of the $j$-th particle
at time $t$. If the deformation is exactly affine,
we simply have $\Delta_{z}(\bm{r},t) = 0$. We calculate the power spectrum
of the non-affine displacement field $\Delta_{z}(\bm{r},t)$ in the Fourier
space, $P(\bm{q})$ ($\bm{q} = 
[ q_{x}, q_{y}, q_{z} ] $ is the wavenumber vector).
From the symmetry of the system, 
we use the cylindrical coordinate
$q_{x} = q_{r} \cos \varphi$ and $q_{y} = q_{r} \sin \varphi$ ($0 \le \varphi < 2 \pi$),
and take the average over $\varphi$. Then we have two-dimensional
scattering power spectrum field $P(q_{r},q_{z})$.
$P(q_{r}) = P(q_{r},q_{z} = 0)$ characterizes the non-affine deformation in the $xy$-plane.
Figure~\ref{nonaffine_deformation_spectra_0_90} shows the
one-dimensional power spectra for $\theta = 0^{\circ}$ and $90^{\circ}$ at $\varepsilon = 1$,
with various strain rates $\dot{\varepsilon}$.
The peak wavenumbers $q^{*}$ in Figure~\ref{nonaffine_deformation_spectra_0_90}(a) 
correspond to the characteristic wavenumbers of undulated layers by the buckling.
We observe that the peak wavenumber decreases as $\dot{\varepsilon}$ decreases
($q^{*} = 0.20$ for $\dot{\varepsilon} \le 0.002$ and
$q^{*} = 0.39$ for $\dot{\varepsilon} \ge 0.005$).
We also observe that the peak value $P(q^{*})$ increases as $\dot{\varepsilon}$ decreases.
This means that the non-affine deformations by buckling is enhanced for small $\dot{\varepsilon}$.
We observe similar trends in the power spectra for $\theta = 90^{\circ}$ (Figure~\ref{nonaffine_deformation_spectra_0_90}(b)).
Therefore, the characteristic magnitude and length scale of the non-affine deformation 
increase as the strain rate decreases.
Although the characteristic length scale depends on the strain rate,
non-affine deformations are commonly observed for all the cases.
We consider that the structural deformation is qualitatively independent
of the strain rate.


\begin{figure}[tb!]
 \centering
{\includegraphics[width=0.5\linewidth,clip]{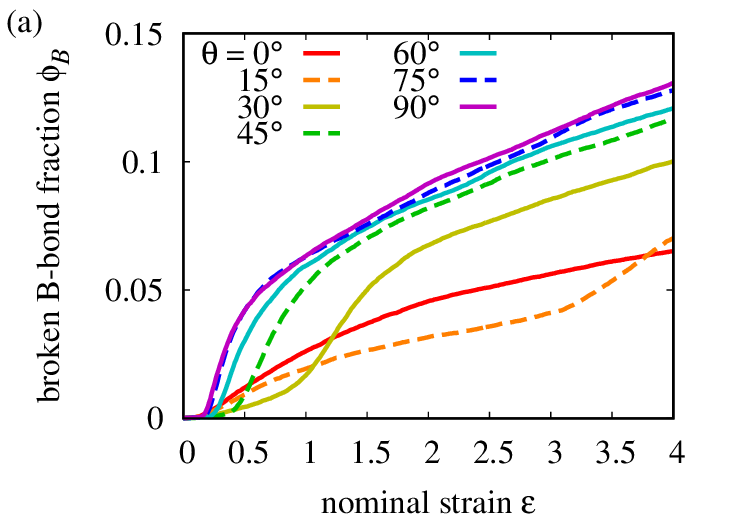}%
\includegraphics[width=0.5\linewidth,clip]{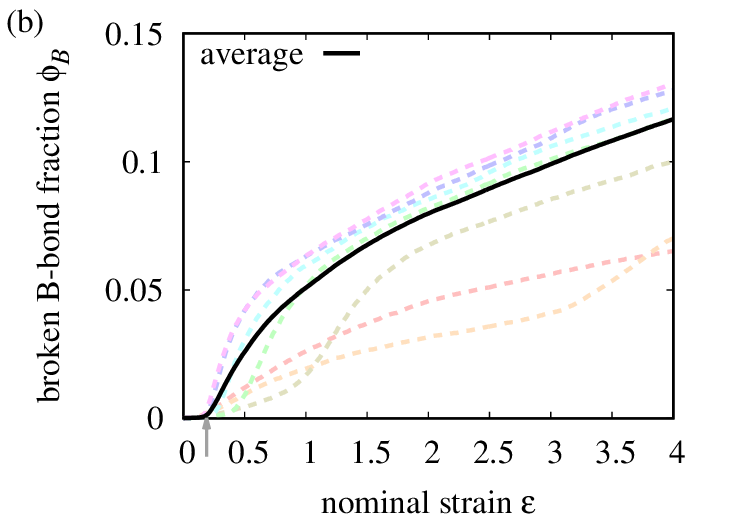}}
 \caption{Broken B-bond fractions during elongation.
 (a) Broken B-bond fractions with different lamellar stacking directions.
 The nominal strain rate is $\dot{\varepsilon} = 0.002$.
 (b) Average broken B-bond fraction calculated from data for various
 lamellar stacking directions (shown in (a)). For comparison,
 data for different lamellar stacking directions
 are shown as dotted curves. The gray arrow indicates the
 characteristic strain where B-bonds start to break, $\varepsilon_{b}' = 0.19$.
}
 \label{bond_fraction_different_directions_and_average}
\end{figure}

To study how the breakage of B-bonds is related to the
structural change, we calculate the fraction of broken B-bonds $\phi_{B}$
as a function of $\varepsilon$. By comparing $\phi_{B}$ with
the structural change or mechanical properties, we can discuss
whether the breakage is important or not.
Figure~\ref{bond_fraction_different_directions_and_average}(a)
shows the broken B-bond fractions for different directions
with $\dot{\varepsilon} = 0.002$.
($\phi_{B}$ is not that sensitive to $\dot{\varepsilon}$.)
From Figure~\ref{bond_fraction_different_directions_and_average}(a),
we observe that $\phi_{B}$ strongly depends on the lamellar stacking
direction. This is consistent with the snapshots in Figure~\ref{snapshots_different_directions}.
Interestingly, we observe that $\phi_{B} \approx 0$ for all $\theta$,
if the strain is sufficiently small.
This means that the deformation is reversible in this region.
But as the strain increases, $\phi_{B}$ rapidly increase.
The macroscopic properties should be determined by the average
value of $\phi_{B}$ rather than the value for a specific $\phi_{B}$.
This is because a macroscopic specimen contain spherulites,
and spherulites consist of crystalline lamellar structures with
various stacking directions.
Thus we calculate the average under the assumption that
lamellar stacking direction is totally random in a macroscopic specimen.
(See Appendix~\ref{average_over_lamellar_stacking_directions}.)
Figure~\ref{bond_fraction_different_directions_and_average}(b)
shows the thus calculated average broken B-bond fraction.
The characteristic strain where B-bonds start to break is estimated
as $\varepsilon'_{b} = 0.19$.


\subsection{Small Angle Scattering}
\label{small_angle_scattering}

In experiments, the detailed structural change of the lamellar structures
during uniaxial elongation cannot be directly observed in the real space.
One powerful experimental method to study mesoscopic structural change
is the small-angle
X-ray scattering (SAXS). SAXS gives the information on the lamellar structures
in the reciprocal space. Therefore, it would be informative for us to
calculate the small angle scattering patterns from our simulation data.
We use the density distribution of crystalline particles to calculate
the scattering patterns. We discretize space and calculate the density
field for crystalline particles. (Because the simulation box is a cuboid,
the number of divisions should be tuned to make the mesh size approximately
constant. We set the average mesh size to be $0.25$.)
Then the density field is Fourier-transformed and the scattering
intensity $I(\bm{q})$ 
is calculated as the squared magnitude of the Fourier-transformed
field. To mimic the experimental condition, we take the average
over the different directions (by the method explained in Appendix~\ref{average_over_lamellar_stacking_directions}).
Also, the average over the azimuthal angle in the cylindrical coordinate
is taken. Finally we have two-dimensional
scattering intensity data $I(q_{r},q_{z})$.

\begin{figure}[tb!]
 \centering
{\includegraphics[width=1.0\linewidth,clip]{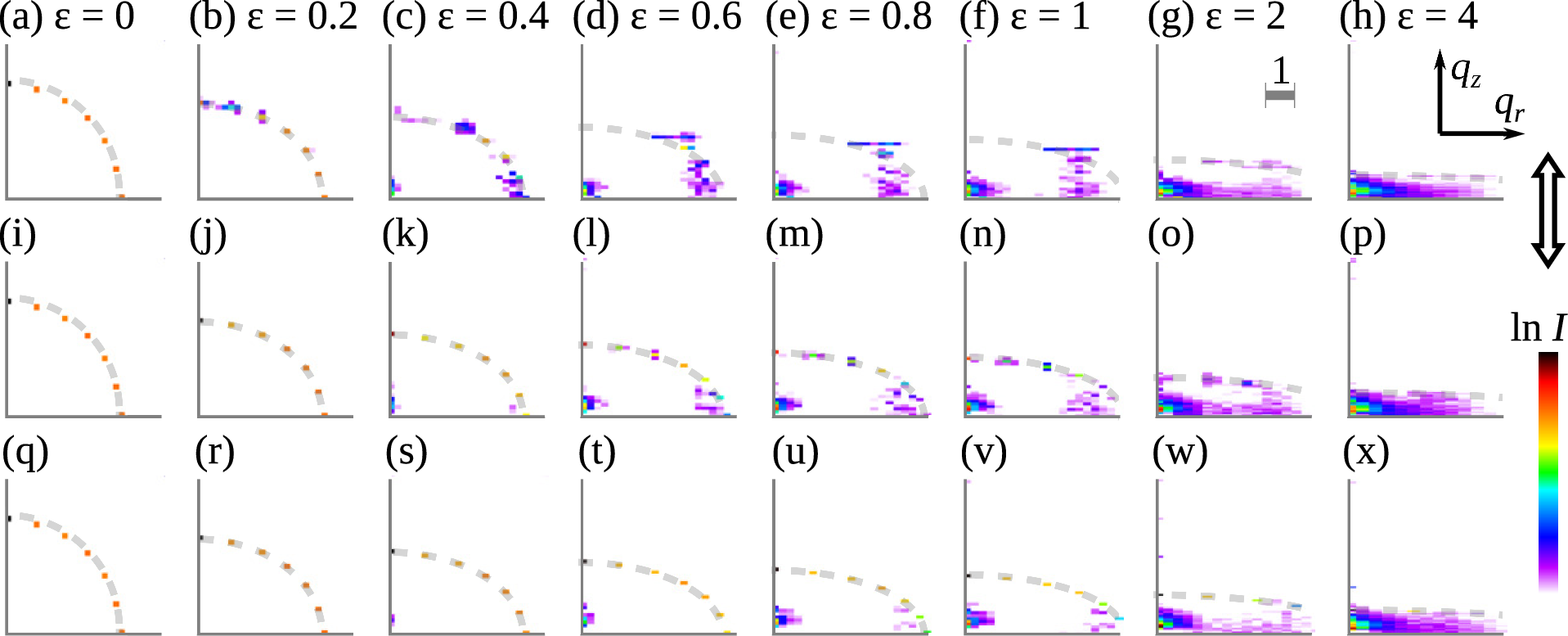}}
 \caption{Two-dimensional small-angle scattering patterns $I(q_{z},q_{r}) $for crystalline
 particles with different nominal strain rates, calculated from snapshots.
 (From the symmetry, only the data for $q_{r} \ge 0$ and $q_{z} \ge 0$ are shown.)
 (a)-(h) $\dot{\varepsilon} = 0.0002$,
 (i)-(p) $\dot{\varepsilon} = 0.002$,
 and (q)-(x) $\dot{\varepsilon} = 0.02$.
 The scattering patterns at $\varepsilon = 0, 0.2, 0.4, 0.6, 0.8, 1, 2,$
 and $4$ are shown.
 The elongation direction is indicated by a white double arrow.
 A gray line segment shows a scale bar.
 Gray dashed curves show scattering patterns for affinely deformed systems.
}
 \label{scattering_patterns_rate_dependence}
\end{figure}

Figure~\ref{scattering_patterns_rate_dependence}
shows the two-dimensional scattering data $I(q_{r},q_{z})$
for $\dot{\varepsilon} = 0.0002, 0.002,$ and $0.02$. 
From the symmetry, only the data for $q_{r} \ge 0$ and $q_{z} \ge 0$ are shown.
Although the resolution of scattering data is rather low, we can observe
some characteristic patterns. Figure~\ref{scattering_patterns_rate_dependence}
will be sufficient to study deformation behaviors qualitatively.
When the strain is sufficiently small ($\varepsilon \lesssim \varepsilon_{b}' = 0.19$),
we observe spots on slightly deformed arcs.
The wavenumber for these spots is consistent with that for crystalline lamellar structures, $q^{*} \approx 2 \pi / D \approx 3.9$.
Different spots correspond to
different lamellar stacking directions. (If we increase the number of
directions, the patterns will approach to a continuum arc.)
We interpret these patterns as affinely deformed lamellar structures.
(The scattering patterns of layer structures under affine
deformations are shown in Figure~\ref{scattering_patterns_rate_dependence}.)
As the strain increases, we observe that the pattern deviates from that
for the simple affine deformation. This behavior is especially clear when
the strain rate is low ($\dot{\epsilon} = 0.0002$).
Some spots disappear and some new spots appear
at the low angle region (Figure~\ref{scattering_patterns_rate_dependence}(c)-(f)).
In addition, the remaining peaks become broader.
These patterns can be interpreted as the formation of large-scale structures
as observed in Figures~\ref{snapshots_different_directions}-\ref{snapshots_rate_dependence_90}.
In the case of high strain rate ($\dot{\varepsilon} = 0.02$), the
arc survives even at $\varepsilon \gtrsim \varepsilon_{b}' = 0.19$ (Figure~\ref{scattering_patterns_rate_dependence}(t)-(v)).
This means that the deformation largely deviates from the affine deformation
as the strain rate decreases. At the high strain regions, diffuse peaks at the
low angle region becomes dominant.
The two-dimensional scattering patterns for $\varepsilon \lesssim 1$
with $\dot{\varepsilon} = 0.0002$
(Figure~\ref{scattering_patterns_rate_dependence}(a)-(f))
seem to be similar to experimental data for low crystallinity polymers,
such as linear low density polyethylene (LLDPE)\cite{Kishimoto-Mita-Ogawa-Takenaka-2020}.

\begin{figure}[tb!]
 \centering
{\includegraphics[width=0.5\linewidth,clip]{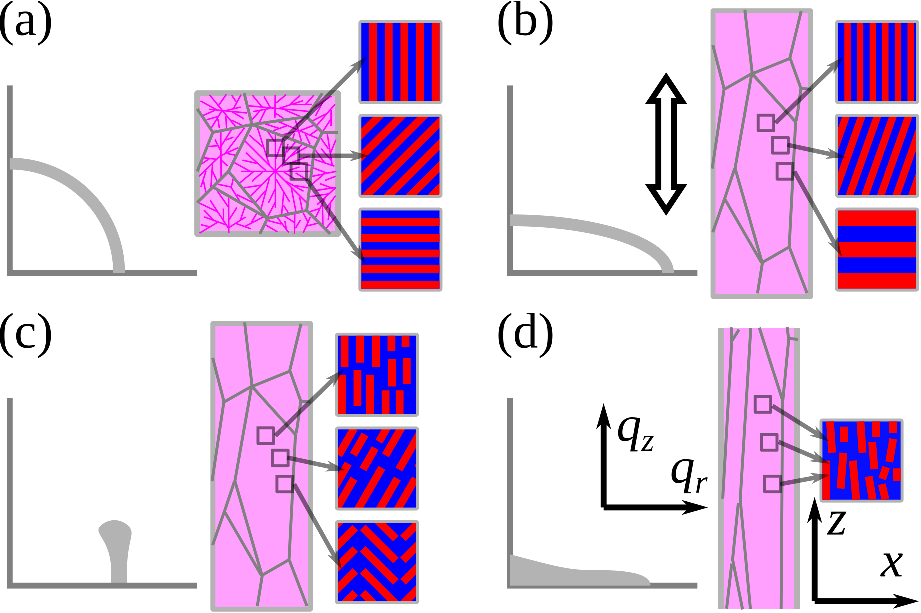}}
 \caption{Schematic images of scattering patterns and structural deformation
 behaviors. (a) An initial isotropic structure. Purple polygons represent
 spherulites. Red and blue regions represent crystalline and amorphous
 regions, respectively.
 The scattering pattern becomes a circle.
 (b) An affinely deformed structure. The scattering pattern becomes an ellipse.
 (c) Non-affinely deformed structure at the medium strain ($\varepsilon \approx 1$).
 Crystalline lamellar structures are buckled or tore. The scattering pattern
 deviates from the ellipse and become a bar-like pattern.
 (d) Non-affinely deformed structure at the high strain ($\varepsilon \approx 4$).
 Crystalline lamellar structures are tore into pieces and align.
 The scattering patterns become a streak-like pattern.
 }
 \label{scattering_schematic_images}
\end{figure}

We show the schematic images of the structural deformation behaviors and corresponding
scattering patterns in Figure~\ref{scattering_schematic_images}.
A macroscopic specimen consists of spherulites, and a spherulite
contains crystalline lamellar structures with various stacking directions. 
If the structure is affinely deformed, we observe that a circular scattering
pattern changes to an elliptic scattering pattern.
The simulation results show that the deformation is non-affine.
At the medium strain ($\varepsilon \approx 1$), crystalline layers are buckled or tore by the breakage
of B-bonds. Buckled layers give diffuse spot patterns\cite{Pope-Keller-1975}.
The tore layers give similar scattering patterns with the affinely deformed
layers. Then the scattering pattern becomes a bar-like pattern.
At the high strain ($\varepsilon \approx 4$), crystalline layers are totally tore into pieces,
and broken pieces align. The scattering pattern becomes streak-like.

\subsection{Stress-Strain Curves}
\label{stress_strain_curves}

\begin{figure}[tb!]
 \centering
{\includegraphics[width=0.5\linewidth,clip]{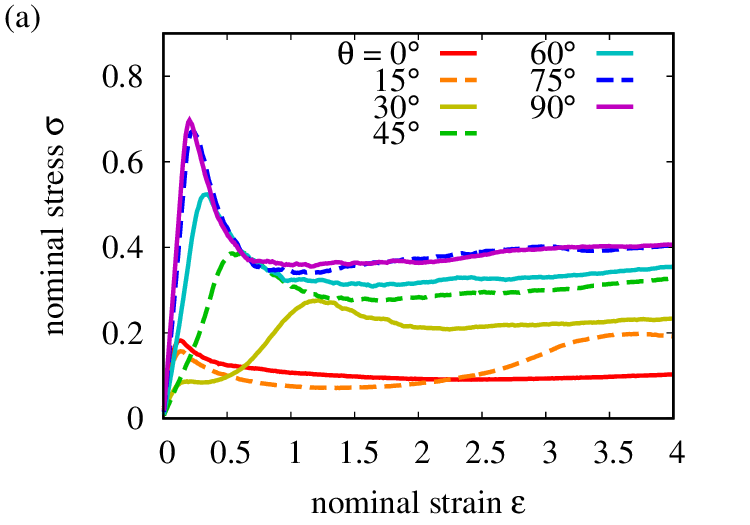}%
\includegraphics[width=0.5\linewidth,clip]{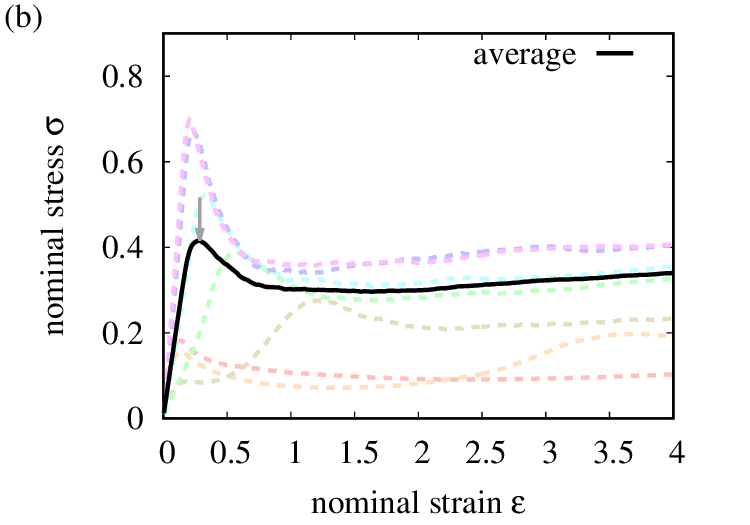}}
 \caption{Stress-strain curves by the coarse-grained crystalline polymer model.
 The nominal strain rate is $\dot{\varepsilon} = 0.002$.
 (a) Nominal stress-strain curves with different lamellar stacking directions $\theta$.
 (b) Average stress-strain curves calculated from data for various
 lamellar stacking directions (shown in (a)). For comparison, data for different lamellar stacking directions
 are shown as dotted curves. The gray arrow indicates the yield point
 (which is determined as the local maximum).
}
 \label{stress_strain_different_directions_and_average}
\end{figure}

Figure~\ref{stress_strain_different_directions_and_average}(a)
shows typical stress-strain curves for $\dot{\varepsilon} = 0.002$
with different lamellar stacking directions $\theta$.
We observe that the stress level strongly depends on $\theta$.
Roughly speaking, the stress increases as $\theta$ increases.
The difference of the stress levels can be understood as follows.
If $\theta$ is large, crystalline layers are directly stretched
and B-bonds are stretched. Then B-bond stretch potential generates high stress.
On the other hand, if $\theta$ is small,
crystalline layers are compressed and bend (by the buckling).
The stress generated
by bending and tilt potentials is smaller than that by B-bond stretch potential.

In all the cases, we observe clear yield behaviors.
For $\theta = 0^{\circ}, 75^{\circ}$ and $90^{\circ}$, the stress-strain curves
have peaks at relatively small $\varepsilon$. The shapes of
stress-strain curves are similar to typical experimental data\cite{Strobl-book,Young-Lovell-book}
(see Figure~\ref{stress_strain_hdpe}),
but judging from structures, the yield mechanism is not unique.
In the case of $\theta = 75^{\circ}$ and $90^{\circ}$, brittle
crystalline layers are stretched and broken into small pieces.
In the case of $\theta = 0^{\circ}$, the compressed crystalline
layers exhibit buckling and then broken.
For $\theta = 45^{\circ}$ and $60^{\circ}$, the stress-strain
curves have similar peaks but at larger strains.
This will be due to the rotation of lamellar structures.
Lamellar structures are rotated by elongation, and then crystalline
layers are broken into pieces when
the stacking direction becomes perpendicular to the elongation direction.
For $\theta = 15^{\circ}$ and $30^{\circ}$, we have two peaks.
Two peaks would be related to different yield mechanisms.
The first peak can be attributed the yield by buckling, and the second peak
can be attributed to the yield by breakage of crystalline layers after rotation.
The shapes of these stress-strain curves seem to be much different
from experimental data.
The stress levels for $\theta \le 30^{\circ}$ is lower than
those of $\theta \ge 45^{\circ}$. For $\theta \ge 45^{\circ}$, the
yield is caused by the tearing mode. On the other hand,
for $\theta \le 30^{\circ}$, the yield is caused by the buckling mode.
The difference of the stress levels will be attributed to the difference
of the deformation modes.

As we explained, in a macroscopic specimen, lamellar stacking directions are not
homogeneous. Thus, to compare the simulation results with
the data for a macroscopic specimen, we should take the average
over different lamellar stacking directions. In this work,
we simply take the weighted average of stress-strain curves
to estimate macroscopic average stress-strain curve,
by using the methods shown in Appendix~\ref{average_over_lamellar_stacking_directions}.
(In reality, the spherulite structures and mechanical balance
between different parts should be considered. However, it would
be a very difficult task and we ignore them for simplicity.)
Figure~\ref{stress_strain_different_directions_and_average}(b)
shows the average stress-strain curve calculated from
data in Figure~\ref{stress_strain_different_directions_and_average}(a).
We observe that the clear yield behavior is observed for the
average stress-strain curve. Moreover, yield behaviors observed
at relatively large strains (in Figure~\ref{stress_strain_different_directions_and_average}(a)) are not observed.
Instead, the stress level becomes insensitive to the strain at the high
strain region. The overall shape of the average stress-strain
curve is very close to typical experimental data such as Figure~\ref{stress_strain_hdpe}.
The yield strain and stress ($\varepsilon_{y}$ and $\sigma_{y}$)
can be estimated as the strain and stress at the local maximum: $\varepsilon_{y} = 0.29$ and $\sigma_{y} = 0.42$.
The Young's modulus $E$ can be estimated by fitting the data for the low
strain region ($0 \le \varepsilon \le 0.075$) to $\sigma = E \varepsilon$.
The fitting gives $E = 2.09$.
The yield strain $\varepsilon_{y} = 0.29$
is slightly higher than the characteristic strain
where the B-bond start to break ($\varepsilon_{b}' = 0.19$).
The macroscopic yield seems to be triggered
by breakage of B-bonds.
We consider that our coarse-grained model reasonably
reproduces both the structural change and mechanical properties
of crystalline polymer solids.

\begin{figure}[tb!]
 \centering
{\includegraphics[width=0.5\linewidth,clip]{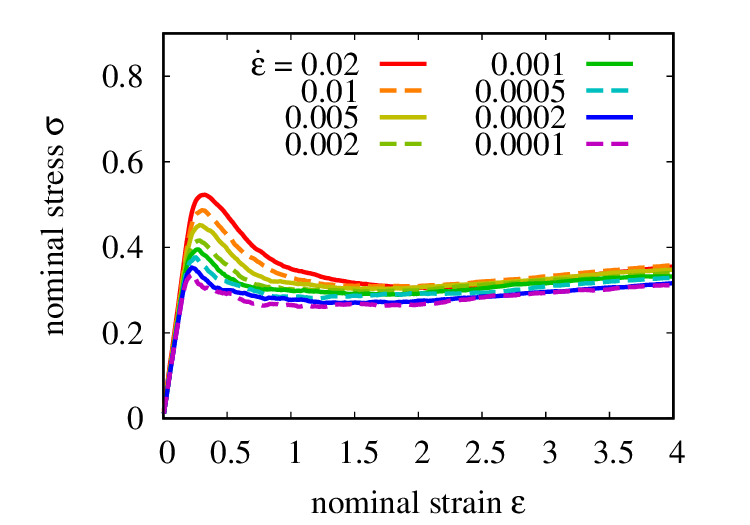}}
 \caption{Stress-strain curves averaged over lamellar stacking directions with
 various nominal strain rates $\dot{\varepsilon}$.
}
 \label{stress_strain_rate_dependence} 
\end{figure}

\begin{figure}[tb!]
 \centering
{\includegraphics[width=0.5\linewidth,clip]{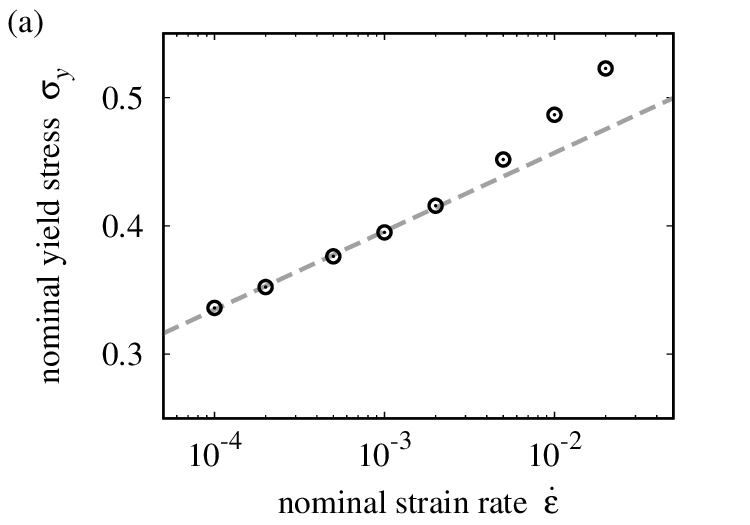}%
\includegraphics[width=0.5\linewidth,clip]{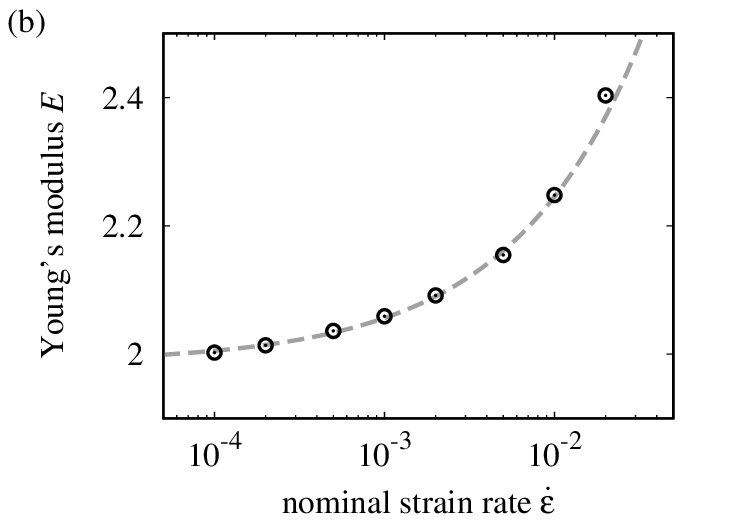}}
 \caption{Strain rate dependence of (a) the yield stress $\sigma_{y}$
 and (b) the Young's modulus $E$.
 Symbols show the simulation data
 and the gray dashed curves are the fitting results.
}
 \label{yield_stress_young_modulus_rate_dependence}
\end{figure}

In experiments, the yield stress is known to depend on the strain rate.
The strain rate dependence is known to be described by the Eyring type relation.
Figure~\ref{stress_strain_rate_dependence} shows average stress-strain
curves with different strain rates. It is clear that the stress-strain curves
depend on the strain rate rather strongly. To study the strain rate dependence
of mechanical properties, we estimate the Young's modulus $E$ and
yield stress $\sigma_{y}$ from the data in Figure~\ref{stress_strain_rate_dependence}.
As explained, the Young's moduli are estimated from the stress-strain data for 
$0 \le \varepsilon \le 0.075$ and the yield stresses are estimated from
the local maxima.
Figure~\ref{yield_stress_young_modulus_rate_dependence}
shows the $\dot{\varepsilon}$ dependence of $\sigma_{y}$ and $E$.
According to the Eyring theory, if the yield stress is relatively large,
the strain rate obeys\cite{Strobl-book,Young-Lovell-book,Desari-Misra-2003}
\begin{equation}
 \label{eyring_relation}
  \dot{\varepsilon} = \frac{\dot{\varepsilon}_{0}}{2}
  \exp\left( \frac{\sigma_{y} v_{a}}{2 k_{B} T}\right).
\end{equation}
Here, $\dot{\varepsilon}_{0}$ and $v_{a}$ are constants.
$\dot{\varepsilon}_{0}$ can be interpreted as the rate constant and
$v_{a}$ is so-called the activation volume.
From eq~\eqref{eyring_relation}, $\sigma_{y}$ depends on $\dot{\varepsilon}$ as
$\sigma_{y} \propto \ln \dot{\varepsilon}$.
The yield stress data in Figure~\ref{yield_stress_young_modulus_rate_dependence}(a)
obeys the Eyring type relation except some points at the large strain rate region.
If we use the data points for $\dot{\varepsilon} \le 0.002$,
the simulation data can be fitted well to eq~\eqref{eyring_relation}.
We have $v_{a} / T_{\text{eff}} = 37.7$
($v_{a} = 0.377$)
and $\dot{\varepsilon}_{0} = 6.7 \times 10^{-10}$.
The Young's modulus data in Figure~\ref{yield_stress_young_modulus_rate_dependence}(b)
can be fitted well to the following empirical relation:
\begin{equation}
 \label{empirical_relation_young_modulus}
  E = E_{0} (1 + c \dot{\varepsilon}^{\alpha}),
\end{equation}
where $E_{0}$ is the modulus at the zero strain rate limit,
and $c$ and $\alpha$ are constants.
By fitting the data points for $\dot{\varepsilon} \le 0.01$ to eq~\eqref{empirical_relation_young_modulus},
we have $E_{0} = 1.99$, $c = 1.8$, and $\alpha = 0.57$.
From the fact that the Eyring type strain rate dependence for the yield
stress is observed in our coarse-grained model, we consider that our
model reasonably reproduce yield behaviors of crystalline polymer solids.

\section{Discussions}
\label{discussions}

\subsection{Comparison with Elastic Network}
\label{comparison_with_elastic_network}

Our model can reproduce the yield behaviors associated with the
breakage of B-bonds.
From Figure~\ref{bond_fraction_different_directions_and_average},
we find that  the broken B-bond fraction increases even after the yield point.
Thus, although the stress is almost unchanged after the yield, 
crystalline layers are gradually damaged. This means that
the motion of broken pieces of crystalline layers is rather
strongly constrained. This implies that the stress after the
yield reflects frustrated structures formed by the broken crystalline
layers. The broken crystalline layers may act as jammed particles\cite{vanHecke-2009,Behringer-Chakraborty-2019}
which generate high stresses.

\begin{figure}[tb!]
 \centering
{\includegraphics[width=0.5\linewidth,clip]{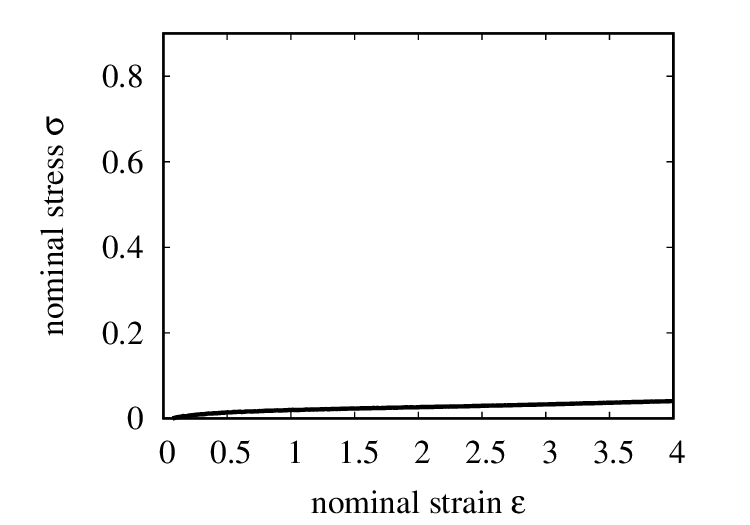}}
 \caption{A stress-strain curve for an elastic network formed only
 by D-bonds. The nominal strain rate is $\dot{\varepsilon} = 0.002$.}
 \label{stress_strain_elastic_network} 
\end{figure}

If there is no frustration and the broken crystalline layers
just work as crosslinks, the resulting stress would be comparable
to that of an elastic network formed only by D-bonds.
Here we perform an elongation simulation for a network
without any B-bonds, for comparison. The simulation conditions
are the same as those for the data in Figure~\ref{stress_strain_different_directions_and_average},
except that B-bonds are not formed at all.
Figure~\ref{stress_strain_elastic_network} 
shows a stress-strain curve for an elastic network only by D-bonds.
(The strain rate is $\dot{\varepsilon} = 0.002$.
The stress-strain curve of an elastic network is almost independent of the strain rate.)
For a relatively large strain region ($1 \le \varepsilon \le 4$),
the stress-strain curve in Figure~\ref{stress_strain_elastic_network} can
be fitted well to the following form: $\sigma = G \epsilon + \sigma_{0}$
($G$ and $\sigma_{0}$ are constants). Fitting gives
$G = 6.83 \times 10^{-3}$, and this value can be interpreted as the shear modulus
of the network.
We observe that the stress of the elastic network
and the estimated modulus $G$ are much lower than the stress of
crystalline lamellar structures in Figure~\ref{stress_strain_different_directions_and_average}.
Thus we conclude that our coarse-grained crystalline polymer model
does not behave as a simple elastic network even after the yield.
If we interpret a structure under a large deformation as
composite of solid domains in an elastomer with the shear modulus $G$,
the stress level increases. The high stress level in Figure~\ref{stress_strain_different_directions_and_average}
may be partially attributed to the composite effect.

\subsection{Lamellar Cluster Units}
\label{lamellar_cluster_units}

In Figures~\ref{snapshots_different_directions}-\ref{snapshots_rate_dependence_90},
we observed that crystalline layers are broken into pieces by deformation
and broken pieces move non-affinely and collectively.
The collective motion will be due to the existence of the network by D-bonds.
The broken pieces are not isolated but connected by D-bonds, and thus
cannot move freely.
As a result, neighboring pieces move together and
non-affine motions will be observed.

On average, B-bonds start to break when the strain exceeds the
characteristic strain of B-bond breakage, $\varepsilon_{b}' = 0.19$ (Figure~\ref{bond_fraction_different_directions_and_average}(b)).
This characteristic strain is close but slightly lower than
the yield strain $\varepsilon_{y} = 0.29$.
Thus, we consider that  the yield originates from
breakage of B-bonds in our coarse-grained model. The fraction of broken B-bonds is very small around
the yield point. Even under large deformation (such as $\varepsilon = 4$),
the broken B-bond fraction is about $0.1$.
This means that only some limited parts of the crystalline layers are broken, and
remaining parts survive. Actually, in Figure~\ref{snapshots_different_directions},
we observe clear crystalline layers even at $\varepsilon = 4$.

These results are not consistent with some old models,
where deformed crystals are unfolded and then recrystallize\cite{Peterlin-1987}.
Our simulation results support
the lamellar cluster model by Nitta and Takayanagi\cite{Nitta-Takayanagi-2000,Nitta-Takayanagi-2003}.
Nitta and Takayanagi considered the deformation and breakage of
crystalline layers in the crystalline lamellar structures. Crystalline
layers are connected by so-called the tie molecules. 
When the layers stacked in the direction of $\theta = 0^{\circ}$ is
elongated, crystalline layers will be broken into pieces. But these
broken pieces are connected by tie molecules. Some pieces connected
by tie molecules will move collectively, which correspond to
a kinetic unit (the lamellar cluster unit). Thus
crystalline lamellar structures are broken
into lamellar cluster units. The yield and plastic deformations can be
interpreted as the formation of lamellar cluster units and rearrangements
of lamellar cluster units.
To form lamellar cluster units, only a small fraction of crystalline
layers are required to be broken.
This is consistent with our simulation data. Moreover,
collectively moving pieces of crystalline layers observed in our
simulations can be interpreted as a lamellar cluster unit.

However, we should note that
our model cannot be directly analyzed in terms of the lamellar
cluster model. In the lamellar cluster model, the tie molecule
(a polymer chain which connects multiple crystalline layers)
is considered to play an important role\cite{Nitta-Takayanagi-2000,Nitta-Takayanagi-2003}.
The size of a lamellar cluster unit is comparable to
the tie molecule size.
In our model, individual
polymer chains are smeared out and we cannot identify where
tie molecules are.
Also, we observe that the crystalline layers can be broken by the tearing mode (for example at $\theta = 90^{\circ}$), in addition to the buckling mode.
Nonetheless, we observe  collectively
moving units of the broken pieces. Although the detailed
descriptions or assumptions are different, the formation of collectively
moving large structures will be common.

\subsection{Constitutive Instability and Neck Formation}
\label{constitutive_instability_and_neck_formation}

In our simulation model, a simulation box is assumed to be
homogeneously deformed. This means that our model cannot directly
reproduce the neck formation which is commonly observed for
crystalline polymer solids.
Nonetheless, we expect that our model will be useful to
consider neck formation behaviors.
In the stress-strain curves in Figures~\ref{stress_strain_different_directions_and_average}
and \ref{stress_strain_rate_dependence}, we clearly find
the local maximum which are attributed to yield points.
The existence of the local maximum means that our crystalline
polymer model is not stable from the view point of constitutive relation\cite{Strobl-book,Young-Lovell-book,Ericksen-1975}.
The situation is qualitatively similar to shear bands\cite{Fielding-2007,Sato-Yuan-Kawakatsu-2010}.

\begin{figure}[tb!]
 \centering
{\includegraphics[width=0.5\linewidth,clip]{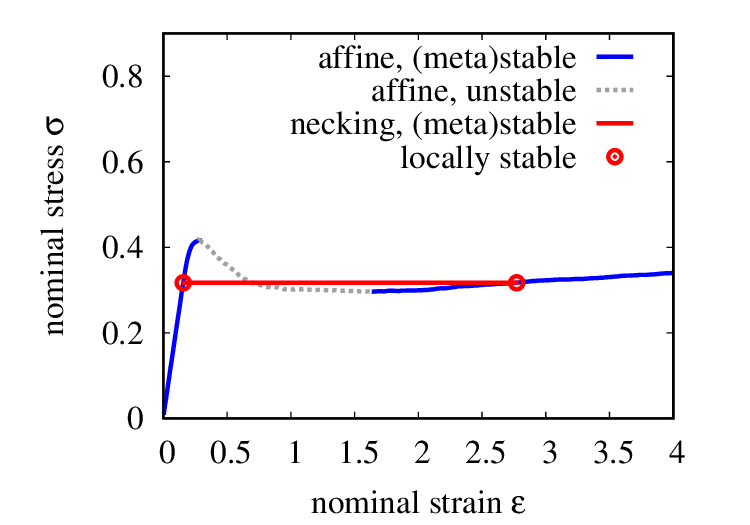}}
 \caption{Maxwell construction for the stress-strain curve in Figure~\ref{stress_strain_different_directions_and_average}(b).
 The blue solid show the (meta)stable region where the macroscopic deformation
 is homogeneous, while the gray dotted curve show the unstable region.
 The red solid line shows the stress for ideally necked states.
 On this line, two states with different nominal strains $\varepsilon_{u} = 0.16$ and $\varepsilon_{n} = 2.77$ (indicated by symbols)
 coexist and the nominal stress is constant: $\sigma_{n} = 0.317$.
}
 \label{maxwell_construction} 
\end{figure}

If $d\sigma / d\varepsilon < 0$, the affine and homogeneous
deformation becomes unstable. Thus, if the specimen is allowed to
deform inhomogeneously, we will observe that the coexistence of
low- and high-strain regions. We can utilize the Maxwell
construction, which is widely used to construct phase diagrams\cite{Strobl-book,Landau-Lifshitz-statistical-book}.
Figure~\ref{maxwell_construction} shows the result of the Maxwell
construction for the stress-strain curve in
Figure~\ref{stress_strain_different_directions_and_average}(b).
Maxwell construction
gives the necking stress $\sigma_{n} = 0.317$ and strains
for necked and unnecked regions $\varepsilon_{n} = 2.77$ and
$\varepsilon_{u} = 0.16$.

The macroscopic mechanical behavior estimated
from Figure~\ref{maxwell_construction} is as follows.
When we deform a macroscopic specimen with a constant nominal
strain rate, it will be deformed affinely
until the yield point.
If the applied strain exceeds the yield strain $\varepsilon_{y} = 0.29$,
homogeneous deformation becomes unstable and shear bands will appear.
Then, shear bands gradually
grow and coexistence of low- and high-strain regions 
($\varepsilon_{u} = 0.16$ and $\varepsilon_{n} = 2.77$) and
the neck propagation fronts between them will be formed.
When the applied strain is further increased, the neck propagation will
occur but the stress is constant ($\sigma_{n} = 0.317$).
This constant stress continues until the applied strain
coincides to $\varepsilon_{n}$. If the applied strain exceeds
$\varepsilon_{n}$, the specimen will be deformed homogeneously again.
The stress starts to increase in this region.

It would be fair to mention that the strain at the high-strain
region, $\varepsilon_{n} = 2.77$, seems to be small compared with typical
experimental data.
(For example, from the stress-strain curve in Figure~\ref{stress_strain_hdpe},
we can estimate $\varepsilon_{n} \approx 6$.)
But this value depends on model parameters such as
potential parameters and the strain rate.
In addition, we ignored mechanical balance between different mesoscopic
regions. Thus we expect that
the value of $\varepsilon_{n}$ may be underestimated.
We believe that the essential physical mechanisms are captured, at least
qualitatively, by our model.

\subsection{Application to Block Copolymers}
\label{application_to_block_copolymers}

The coarse-grained model we constructed in this work may be
used to study structural deformation and mechanical properties of
other polymeric systems. In our coarse-grained model, the microscopic
crystalline structures are not explicitly considered. Thus,
as long as there are two phases and one phase is brittle, our
model can be utilized.

A simple yet interesting application is mechanical properties of
some block copolymers\cite{Adhikari-Michler-2004}
The equilibrium microphase separation structures as well
as microphase separation dynamics have been extensively studied by
continuum field models\cite{Fredrickson-book,Leibler-1980,Ohta-Kawasaki-1986,Matsen-Bates-1996,Muller-Schmid-2005,Uneyama-Doi-2005}.
The mechanical properties reflect microphase separation structures.

For example, ABA triblock copolymers form microphase
separation structures such as cylinder and lamellar structures.
If A block is glassy and B block is rubbery, then the A-rich
region is hard but brittle while the B-rich region
is soft but ductile.
This is qualitatively very similar
to crystalline polymer solids.
Although mechanical properties of ABA triblock copolymers have
been studied with coarse-grained molecular dynamics simulations\cite{Aoyagi-Honda-Doi-2002,Makke-Perez-Lame-Barrat-2012,Makke-Lame-Perez-Barrat-2012,Makke-Lame-Perez-Barrat-2013,Hagita-Akutagawa-Tominaga-Jinnai-2019},
the characteristic time scale of coarse-grained MD is short.
(This situation will be similar to the case of MD simulations
for crystalline polymers.)
To overcome this problem, our
coarse-grained model will be useful.
From the view point of highly coarse-grained models, whether the
hard and brittle regions are formed by crystals or glasses are not
important. By tuning potential parameters and initial structures,
various microphase separation structures will be handled by our
coarse-grained model.

\subsection{Buckling of Layers}
\label{buckling_of_layers}

As discussed in Sec.~\ref{application_to_block_copolymers},
microphase separation structures by ABA triblock copolymers exhibit
similar mechanical behaviors to crystalline polymer solids.
Makke et al\cite{Makke-Lame-Perez-Barrat-2013} studied the buckling
of block copolymer lamellar structures by coarse-grained MD simulations.
They showed that the buckling behavior depends on the strain rate.
When the strain rate is low, the buckling can be explained by the
elastic buckling of glassy layers. (The characteristic wavenumber of
the buckling should be consistent with the system size, and thus the
observed buckling behavior depends on the system size.)
On the other hand, when the strain rate is high, the voids (or cavities)
are formed during the elongation, and the void formation changes
the structural deformation behavior.

In our simulation model, the void formation is not considered.
Thus the competition between the buckling and the void formation does
not occur in our simulations. The strain rate dependence of the
buckling behavior in our simulations (Figures~\ref{snapshots_different_directions}, \ref{snapshots_rate_dependence_0},
\ref{nonaffine_deformation_spectra_0_90})
is considered to be due to a different mechanism.
We observe that the characteristic wavenumber decreases as the
strain rate decreases (Figure~\ref{nonaffine_deformation_spectra_0_90}). In the low strain rate cases, the characteristic
wavenumber seems to be limited by the system size. The situation
is similar to that reported by Makke et al.
In the high strain rate cases, the wavenumber is rather large and
the magnitude is relatively low.

These strain rate dependent buckling behaviors may be understood
as follows. The buckling mechanism itself is essentially the same
as that proposed by Makke et al~\cite{Makke-Lame-Perez-Barrat-2013}.
The characteristic wavenumber is determined by the elastic properties of a
crystalline layer. If the system size is not sufficiently large,
the buckling predicted by the linear elasticity model will occur.
(If the system size is not sufficiently large, the characteristic wavenumber 
will be suppressed to be compatible to the
system size.)
When the deformation mode with the characteristic wavenumber
becomes unstable, the magnitude of that deformation starts to grow.
However, to complete such large-scale deformations, relatively long time
will be required. If the strain rate is high, the macroscopic deformation
proceeds before the undulation of the characteristic wavenumber grows.
Then, undulation modes with higher wavenumbers will be activated.
These modes will grow faster than the large wavenumber mode.
As a result, the observed characteristic wavenumber increases.
We consider that the strain rate dependent buckling may be
attributed to the competition between the strain rate and
the characteristic time scale of buckling modes.

\section{Conclusions}
\label{conclusions}

We proposed a highly coarse-grained simulation model for 
crystalline polymer solids. We modeled a crystalline polymer solid
by using coarse-grained particles of which size is comparable to
the crystalline and amorphous layers in a crystalline lamellar structure.
By using such highly coarse-grained particles, our model can
handle large-scale and long-time simulations with small calculation costs
compared with typical MD simulations.
The particles are connected by two types of bonds:
B-bond (brittle or breakable) and D-bond (ductile), to form a
network structure.
The short-range interaction potential between particles is expressed as the Hertz type
contact interaction, and D-bonds are expressed as harmonic potentials.
To express the interaction by B-bonds, we employed directors for
coarse-grained particles. By using relative particle positrons and
directors, the stretching, tilt, and bending potentials for
B-bonds were formulated.
As a dynamics model, simple Langevin equations and
a simple breakage rule for B-bonds were employed.

We performed elongation simulations with the constructed coarse-grained model,
and showed that both the structural changes and mechanical behaviors
can be reasonably reproduced.
Simulations were performed with several different lamellar stacking directions.
To estimate macroscopic properties, we calculated weighted
averages of physical quantities over different lamellar stacking directions.
Macroscopic two-dimensional scattering patterns and
stress-strain curves can be reasonably reproduced in this method.
The structural and mechanical behaviors of our model
are qualitatively consistent with experiments.
If the applied strain is small, B-bonds are not broken and
the structures are deformed affinely. The linear elastic behavior
is observed in this region.
As we increase the strain and if the strain exceeds a characteristic
value, B-bonds start to be broken. As a result, crystalline layers
are broken into pieces. The structures are deformed
non-affinely, and the yield behaviors are observed.
When the strain is further increased, broken pieces of
crystalline layers move collectively, in a similar way to
the lamellar cluster units.
In addition, the yield stress obeys the Eyring type relation.
We expect that our coarse-grained model will be useful to study
structural and mechanical behaviors of various crystalline
polymer solids.
larger-scale simulations will be an interesting future work.
We may be able to the buckling behavior and scattering patterns
in detail, with larger-scale simulations.

\section*{Acknowledgment}

This work was supported by JST PRESTO Grant Number JPMJPR1992, and
Grant-in-Aid (KAKENHI) for Scientific Research Grant B No.~JP23K25839
from Ministry of Education, Culture, Sports, Science, and Technology.

\appendix
\section*{Appendix}

\section{Experimental}
\label{experimental}

In this appendix, we describe details of the experiment for
the stress-strain data shown in Figure~\ref{stress_strain_hdpe}.
High density polyethylene (HDPE, 547999, Sigma-Aldrich) is used as received.
Pellets of HDPE are pressure-molded by a mini test press
(MP-2FH, Toyoseiki).
The pellets are put between two copper plates coated with
$0.05\text{mm}$-thick polyethylene terephthalate films. An aluminum spacer
with $0.5\text{mm}$ thickness is used to control the thickness of
a HDPE sheet. The pellets are pressure-molded 
at $180^{\circ}\text{C}$ and $10\text{MPa}$ for $5 \text{min}$.
The molded HDPE sheet is quenched to $0^{\circ}\text{C}$ by
ice water.
A dumbbell-shaped specimen (the gauge section is $5 \text{mm} \times 20 \text{mm}$)
is punched with a cutter (SDK-4219-19, Dumbbell).
For the thus prepared specimen, a uniaxial elongation test is
performed with a universal testing machine (AGS-X, Shimadzu).
A $500 \text{N}$ load cell is used, and a contact-type extensometer
(DSES-1000, Shimadzu) with a $20 \text{mm}$ gauge bar is
used to measure the actual elongation of the gauge section.
The elongation speed is $20 \text{mm}/\text{min}$, 
and the corresponding nominal strain rate is $1.6 \times 10^{-2} \text{s}^{-1}$.
The test is conducted at the room temperature.
From the measured stress-strain curve, several mechanical parameters
are estimated. The Young's modulus is
estimated as $6.07 \times 10^{2} \text{MPa}$ (from data points
for $\varepsilon < 0.01$), the yield strain
and stress are estimated as $\varepsilon_{y} = 0.14$ and $\sigma_{y} = 15.4\text{MPa}$
(from the local maximum),
and the strain and stress at break are estimated as
$\varepsilon_{b} = 10.7$ and $\sigma_{b} = 28.0\text{MPa}$.

\section{Calculations for Force, Torque, and Stress}
\label{calculations_for_force_torque_and_stress}

In this appendix, we calculate forces and torques by interaction potentials
\eqref{contact_potential}-\eqref{bend_potential}, and contributions of
them to the stress tensor.
The position and director of a particle are driven by the force and
torque acting on them,
according to the Langevin equations \eqref{langevin_equation_r} and
\eqref{langevin_equation_u}.
The force and torque can be decomposed into individual potentials
used in eq~\eqref{total_potential}. In a similar way, the stress
can be decomposed into contributions from individual potentials.

All the interaction potential models in our model are
pairwise, and thus all the forces are also pairwise.
For example, for the contact potential $U_{\text{contact}}(\bm{r})$,
the force by this potential is $\bm{F}_{\text{contact}}(\bm{r}) = - \partial U_{\text{contact}}(\bm{r}) / \partial \bm{r}$.
From eqs~\eqref{contact_potential}-\eqref{tilt_potential}, the forces by the contact, D-bond, B-bond and tilt potentials are
calculated as follows:
\begin{align}
 \bm{F}_{\text{contact}}(\bm{r}) & =
  \begin{cases}
   \displaystyle \frac{5 \epsilon}{2 \sigma_{0}}
   \left( 1 - \frac{|\bm{r}|}{\sigma_{0}} \right)^{3/2} \frac{\bm{r}}{|\bm{r}|} & (|\bm{r}| < \sigma_{0}), \\
   0 & (|\bm{r}| \ge \sigma_{0}) ,
  \end{cases} \\
 \bm{F}_{\text{D-bond}}(\bm{r}) 
 & 
 = - k_{\text{D-bond}} \bm{r} , \\
 \bm{F}_{\text{B-bond}}(\bm{r}) 
 & 
 = - k_{\text{B-bond}} (|\bm{r}| - b_{0}) \frac{\bm{r}}{|\bm{r}|} , \\
 \bm{F}_{\text{tilt}}(\bm{r},\bm{u},\bm{u}') 
 & = - k_{\text{tilt}} 
 \left[ \frac{\bm{u} \cdot \bm{r}}{\bm{r}^{2}}  \bm{u}
 + \frac{\bm{u}' \cdot \bm{r}}{\bm{r}^{2}} \bm{u}'
 - \left[
    \left( \frac{\bm{u} \cdot \bm{r}}{\bm{r}^{2}}  \right)^{2}
    + \left( \frac{\bm{u}' \cdot \bm{r}}{\bm{r}^{2}}  \right)^{2}
\right] \bm{r} \right].
\end{align}
The bending potential \eqref{bend_potential} does not depend on the position of the particle and
thus does not contribute to the force acting on the position.
The torque acting on the director can be decomposed into two contributions.
The contributions by the tilt and bending potentials are:
\begin{align}
 \bm{T}_{\text{tilt}}(\bm{r},\bm{u},\bm{u}')
 & = - \frac{\partial U_{\text{tilt}}(\bm{r},\bm{u},\bm{u}')}{\partial \bm{u}} 
 = - k_{\text{tilt}} 
  \frac{(\bm{u} \cdot \bm{r}) \bm{r}}{\bm{r}^{2}},  \\
 \bm{T}_{\text{bend}}(\bm{u},\bm{u}') & = 
 -\frac{\partial U_{\text{bend}}(\bm{u},\bm{u}')}{\partial \bm{u}}
  =  k_{\text{bend}}  (\bm{u} \cdot \bm{u}') \bm{u}' .
\end{align}
Other potentials do not contribute to the force acting on the director.

Next, we calculate the contributions of interaction potentials to
the stress tensor. To do so, we introduce a small virtual deformation:
$\bm{r} \to \bm{r} + \bm{E} \cdot \bm{r}$, 
$\bm{u} \to \bm{u} + (\bm{E} + \bm{E}^{\mathrm{T}}) \cdot \bm{u} / 2$,
and
$\bm{u}' \to \bm{u}' + (\bm{E} + \bm{E}^{\mathrm{T}}) \cdot \bm{u}' / 2$,
as explained in the main text.
Then, the changes of the potential energies
\eqref{contact_potential}-\eqref{tilt_potential} upto the first order in $\bm{E}$ become as follows:
\begin{align}
  V \bm{\sigma}_{\text{contact}}(\bm{r}) : \bm{E} 
 & =   \begin{cases}
   \displaystyle  \frac{5\epsilon}{2 \sigma_{0}} \left( 1 - \frac{|\bm{r}|}{\sigma_{0}} \right)^{3/2} \frac{\bm{r}\bm{r}}{|\bm{r}|} : \bm{E} & (|\bm{r}| < \sigma_{0}), \\
   0 & (|\bm{r}| \ge \sigma_{0}) ,
     \end{cases} \\
 V \bm{\sigma}_{\text{D-bond}}(\bm{r}) : \bm{E} 
 & = k_{\text{D-bond}} \bm{r} \bm{r} : \bm{E}, \\
 V \bm{\sigma}_{\text{B-bond}}(\bm{r}) : \bm{E} 
  & = k_{\text{B-bond}} (|\bm{r}| - b_{0}) \frac{ \bm{r} \bm{r}}{|\bm{r}|} : \bm{E}, \\
\begin{split}
 V \bm{\sigma}_{\text{tilt}}(\bm{r},\bm{u},\bm{u}') : \bm{E} 
  & =
  k_{\text{tilt}}  \bigg[  
  \frac{\bm{u} \cdot \bm{r}}{\bm{r}^{2}}
  \frac{\bm{u} \bm{r} + \bm{r} \bm{u}}{2} 
  + \frac{ \bm{u}' \cdot \bm{r}}{\bm{r}^{2}}
  \frac{\bm{u}' \bm{r} + \bm{r} \bm{u}'}{2} \\
  & \qquad - \left[  \frac{ (\bm{u} \cdot \bm{r})^{2}}{\bm{r}^{2}}
  + \frac{ (\bm{u}' \cdot \bm{r})^{2}}{\bm{r}^{2}}
  \right]   \frac{\bm{r} \bm{r}}{\bm{r}^{2}}
 \bigg] : \bm{E}. 
\end{split}
\end{align}
Thus we have eqs~\eqref{contact_stress}-\eqref{tilt_stress} in the main text.
The bending potential \eqref{bend_potential} does not contribute to the stress.

\section{Estimates for Potential Parameters}
\label{estimates_for_potential_parameters}

In this appendix, we roughly estimate the potential parameters in our coarse-grained model.
We need some elastic constants of crystalline and amorphous phases for the target polymer:
the shear and bulk moduli of the amorphous region, $G_{a}$ and $K_{a}$,
and the shear modulus and Poisson ratio of the crystalline region, $G_{c}$ and $\nu_{c}$.
We also need the crystalline layer thickness $L_{c}$ and the temperature $T$.
We employ following values:
$G_{a} \approx 1 \text{MPa}$ and $K_{a} \approx 1 \text{GPa}$,
$G_{c} \approx 1 \text{GPa}$, $\nu_{c} \approx 0.3 \sim 0.4$,
 $L_{c} \approx 10 \text{nm}$,
and $T \approx 300 \text{K}$.

We consider the parameters for B-bonds.
The potential parameters for B-bonds are $b, k_{\text{B-bond}}$, $k_{\text{tilt}}$, and $k_{\text{bend}}$.
The natural length of the B-bond $b$ should be comparable to the
particle diameter: $b \approx \sigma_{0}$.
Here we consider the
case where the particles form a two-dimensional honeycomb lattice and
interact only via B-bonds. This can be interpreted as a single elastic layer.
If we assume that this layer consists of an isotropic elastic material, we can
calculate its elastic energy. We consider a layer with the thickness $h$.
We express the shear modulus and the Poisson's ratio of the
solid region as $G_{c}$ and $\nu_{c}$. The Young's modulus is $E_{c} = 2 (1 + \nu_{c}) G_{c}$.
If we apply the uniform in-plane shear deformation to the layer, then the
elastic energy per unit area is calculated to be
\begin{equation}
 \label{shear_energy_macro}
 u_{\text{shear}} = \frac{1}{2} G_{c} h \gamma^{2},
\end{equation}
with $\gamma$ being the shear strain. If we slightly bend the layer,
the behavior of the layer would be modeled
as a thin elastic plate\cite{Landau-Lifshitz-elasticity-book}. As a simple bending deformation, we set
two principal curvatures of the plates are as the same constant value.
Then the elastic energy per unit area becomes
\begin{equation}
 \label{bend_energy_macro}
 u_{\text{bend}} = \frac{E_{s} h^{3}}{24 (1 - \nu_{c}^{2})} [(2 C)^{2} + 2 (1 - \nu_{c}) (- C^{2})]
  = \frac{ (1 + \nu_{c})^{2} G_{c} h^{3}}{6 (1 - \nu_{c}^{2})} C^{2}
\end{equation}
with $C$ being the curvature. These elastic energies can be compared with
the potential energy of the model.

We consider the most stable structure as the reference state. If all
the B-bonds have the same size $b$ and they are on the same flat plane,
the B-bond energy is zero.
Also, if all the directors are perpendicular to the layer, the bending
and tilt energies are also zero. The two-dimensional number density of particles is
$2 / \sqrt{3} b^{2}$, and the two-dimensional B-bond density is
$2 \sqrt{3} / b^{2}$. We apply the shear and bending deformation
to this reference state. 

If we apply the in-plane shear deformation, then B-bonds in the layer are deformed.
The bending and tilt potentials are not affected by this shear deformation.
We express the two-dimensional shear deformation as follows:
\begin{equation}
\begin{split}
  \bm{r}' & = 
  \begin{bmatrix}
   \cos \vartheta & - \sin \vartheta \\
   \sin \vartheta & \cos \vartheta
  \end{bmatrix}
  \cdot
  \begin{bmatrix}
   1 & \gamma \\
   0 & 1
  \end{bmatrix}
  \cdot
  \begin{bmatrix}
   \cos \vartheta & \sin \vartheta \\
   - \sin \vartheta & \cos \vartheta
  \end{bmatrix}
  \cdot
  \bm{r} \\
 & =
  \begin{bmatrix}
 1 - \gamma \cos \vartheta \sin \vartheta & \gamma \cos^{2} \vartheta \\
   - \gamma \sin^{2} \vartheta & 1 + \gamma \cos \vartheta \sin \vartheta
  \end{bmatrix}
  \cdot
  \bm{r}  ,
\end{split}
\end{equation}
where $\bm{r}$ and $\bm{r}'$ are the B-bond vectors before and after the deformation,
and $\vartheta$ is the rotation angle.
From the symmetry, we can set $\bm{r} = [b, 0]$ without loss of generality. Then
the B-bond energy is
\begin{equation}
 \begin{split}
  U_{\text{bend}}(\bm{r}') 
  & = 
  \frac{1}{2} k_{\text{B-bond}} 
  \left[\sqrt{b^{2} \left(1 - \gamma \cos \vartheta \sin \vartheta \right) + b^{2} \gamma^{2} \sin^{2} \vartheta} - b_{0} \right]^{2} \\
  & \approx \frac{1}{2} k_{\text{B-bond}} b^{2}
  \gamma^{2} \cos^{2} \vartheta \sin^{2} \vartheta
 \end{split}
\end{equation}
The rotation angle $\vartheta$ is arbitrary and thus we take the average over $\vartheta$.
Then we have the B-bond energy per unit area as
\begin{equation}
 \label{shear_energy_meso}
 u_{\text{shear}} = \frac{2 \sqrt{3}}{b_{0}^{2}} \frac{1}{2 \pi} \int_{0}^{2 \pi} d\vartheta \,
  \frac{1}{2} k_{\text{B-bond}} b_{0}^{2}
  \gamma^{2} \cos^{2} \vartheta \sin^{2} \vartheta
  = \frac{\sqrt{3}}{8}
  k_{\text{B-bond}} \gamma^{2} .
\end{equation}
Comparing eqs \eqref{shear_energy_macro}
and \eqref{shear_energy_meso}, we find
\begin{equation}
 \label{bbond_potential_parameter_estimate}
 k_{\text{B-bond}} \approx \frac{4}{\sqrt{3}} G_{c} h.
\end{equation}

We consider to apply the bending deformation to the layer, by keeping
the B-bond length unchanged. Then the B-bond energy is not affected.
Two angle between neighboring directors $\bm{u}$ and $\bm{u}'$ is estimated as
\begin{equation}
 \varphi \approx C b_{0}.
\end{equation}
Then the bending and tilt energies are
\begin{equation}
 \label{tilt_energy_meso}
 U_{\text{tilt}}(\bm{r},\bm{u},\bm{u}') 
  \approx
 \frac{1}{2} k_{\text{tilt}} \left[ (\sin (\varphi / 2))^{2}  + (\sin (\varphi / 2))^{2} \right]
  \approx 
 \frac{1}{4} k_{\text{tilt}} b_{0}^{2} C^{2} 
\end{equation}
\begin{equation}
 \label{bend_energy_meso}
 U_{\text{bend}}(\bm{u},\bm{u}') 
  \approx
 \frac{1}{2} k_{\text{bend}} \left[ 1 - (\cos \varphi)^{2}  \right]
  \approx 
  \frac{1}{4} k_{\text{bend}} b_{0}^{2} C^{2}
\end{equation}
The potential energy per unit area becomes
\begin{equation}
 \label{total_bend_energy_meso}
 u_{\text{bend}} \approx \frac{2 \sqrt{3}}{b_{0}^{2}} 
  \left(  \frac{1}{4} k_{\text{tilt}} b_{0}^{2} C^{2} 
   + \frac{1}{4} k_{\text{bend}} b_{0}^{2} C^{2}  \right)
  = \frac{\sqrt{3}}{2} (k_{\text{tilt}} + k_{\text{bend}}) C^{2}.
\end{equation}
Then, comparing eqs \eqref{bend_energy_macro} and 
\eqref{total_bend_energy_meso}, we have
\begin{equation}
(k_{\text{tilt}} + k_{\text{bend}}) 
  \approx \frac{(1 + \nu_{s})^{2} G_{c} h^{3}}{3 \sqrt{3} (1 - \nu_{c}^{2})}.
\end{equation}
If we simply set $k_{\text{tilt}} \approx k_{\text{bend}}$, then
\begin{equation}
 \label{tilt_bend_potentials_parameter_estimate_poisson_ratio}
 k_{\text{tilt}} \approx k_{\text{bend}}
  \approx \frac{(1 + \nu_{c}) G_{c} h^{3}}{6 \sqrt{3} (1 - \nu_{c}^{2})}
  \approx \frac{(1 + \nu_{c})^{2}}{24 (1 - \nu_{c}^{2})} k_{\text{B-bond}} h^{2}.
\end{equation}
In the dimensionless unit, we expect that $h \approx \sigma = 1$. 
By substituting  $\nu_{c} \approx 0.3 \sim 0.4$ to eq~\eqref{tilt_bend_potentials_parameter_estimate_poisson_ratio},
we have the following relation for dimensionless potential parameters:
\begin{equation}
 \label{tilt_bend_potentials_parameter_estimate}
 \frac{k_{\text{tilt}}}{ k_{\text{B-bond}}} \approx 
  \frac{k_{\text{bend}}}{ k_{\text{B-bond}}} \approx 0.077 \sim 0.097.
\end{equation}
Eq~\eqref{tilt_bend_potentials_parameter_estimate} means that $k_{\text{tilt}}$ and $k_{\text{bend}}$ are 
about one order smaller than $k_{\text{B-bond}}$.

We consider the potential parameter for D-bonds, $k_{\text{D-bond}}$.
In absence of B-bonds, the particles are connected only by D-bonds and form
an elastic network. The characteristic bond size is roughly estimated as
the particle diameter $\sigma_{0}$. We apply a shear deformation with
the shear strain $\gamma$ to this elastic network. Macroscopically, such
an network can be modeled as an isotropic elastic body.
For a small deformation, the elastic energy per unit volume is estimated to be
\begin{equation}
 \label{shear_energy_amorphous_macro}
 u_{\text{shear}} = \frac{1}{2} G_{a} \gamma^{2}
\end{equation}
where $G_{a}$ is the shear modulus of the amorphous region.
The D-bond potential energy of this network is increased by
the shear deformation.
The energy increase for a D-bond with the bond vector $\bm{r}$ is
\begin{equation}
 U(\bm{r} + \gamma r_{y} \bm{e}_{x}) - U(\bm{r})
  = \frac{1}{2} k_{\text{D-bond}} 
  (2 \gamma r_{x} r_{y} + \gamma^{2} r_{y}^{2}).
\end{equation}
Bond vectors can take any directions and thus we should take the average
over the bond vector to estimate the increase of the potential energy 
per unit volume. If we assume that the bond size is constant, $|\bm{r}| \approx \sigma_{0}$, then
we have
\begin{equation}
 \label{shear_energy_amorphous_meso}
 \begin{split}
  u_{\text{shear}}
  & \approx \nu_{\text{D-bond}} \frac{1}{4 \pi} \int_{0}^{2 \pi} d\vartheta \int_{0}^{\pi} d \varphi \, \sin \varphi \,
 \frac{1}{2} k_{\text{D-bond}} \sigma_{0}^{2}
 (2 \gamma  \cos \vartheta \sin \vartheta \sin^{2} \varphi 
 + \gamma^{2} \sin^{2} \vartheta \sin^{2} \varphi) \\
  & = \frac{1}{6} n_{\text{D-bond}} k_{\text{D-bond}} \sigma_{0}^{2} \gamma^{2}.
 \end{split}
\end{equation}
Here, $n_{\text{D-bond}}$ is the characteristic number density of D-bonds.
From eqs~\eqref{shear_energy_amorphous_macro} and \eqref{shear_energy_amorphous_meso},
the potential parameter is estimated as
\begin{equation}
 k_{\text{D-bond}} \approx \frac{3}{n_{\text{D-bond}} \sigma_{0}^{2}} G_{a}.
\end{equation}
The number density of D-bonds can be related to the average number
of D-bonds per particle $z_{0}$ and the number density of particles $\rho_{0}$,
as $n_{\text{D-bond}} \approx \rho_{0} z_{0} / 2$. The volume of the single particle 
is $\pi \sigma_{0}^{3} / 6$, thus we estimate the typical number density as
$\rho_{0} \approx 6 / \pi \sigma_{0}^{3}$.
We employ the typical
value of the shear modulus of a polymer melt, $G_{a} \approx 1 \text{MPa}$.
The particle size is roughly estimated
as $\sigma_{0} \approx L_{c}$.
By setting $\sigma_{0} \approx 10 \text{nm}$ and
$G_{c} \approx 1 \text{GPa}$,
we have the following estimate for the ratio of $k_{\text{B-bond}}$ to $k_{\text{D-bond}}$:
\begin{equation}
 \label{dbond_potential_parameter_estimate}
 \frac{k_{\text{D-bond}}}{k_{\text{B-bond}}}
  \approx \frac
  {\displaystyle \frac{\pi \sigma_{0}}{z_{0}} \times 1 \text{MPa}}
  {\displaystyle \frac{4\sigma_{0} }{\sqrt{3}} \times 1 \text{GPa} }
  \approx 1.3 \times 10^{-3} z_{0}^{-1}.
\end{equation}
Thus $k_{\text{D-bond}}$ should be much smaller than $k_{\text{B-bond}}$.

We consider the bulk modulus of the amorphous region.
Here we ignore all the contributions of ductile and B-bonds and consider
only the Hertzian contact interaction. We apply the volume strain by
isotropically compressing the system. Then the elastic energy per unit volume
is
\begin{equation}
 \label{compress_energy_macro}
 u_{\text{compress}} = \frac{1}{2} K_{a} \left(\frac{\rho_{0}}{\rho} - 1 \right)^{2}
\end{equation}
where $K_{a}$ is the bulk modulus, and $\rho_{0}$ and $\rho$ are the number densities before and after the compression.
At $\rho_{0} \sigma_{0}^{3} \approx 6 / \pi \approx 1.9$,
the Hertzian spheres form the face-centered cubic (FCC) phase at low temperatures\cite{Pamies-Cacciuto-Frenkel-2009}.
We employ the perfect FCC crystal with the lattice constant $a_{0} = (4 / \rho_{0})^{1/3}$,
in order to estimate the energy increase by volume compression.
We have four particles per unit cell, and one particle is contacted to
$12$ particles. Then, the number of contacts per unit cell is $24$.
The distance between neighboring particles is $|\bm{r}| = a_{0} / \sqrt{2}$. Then
the contact potential is estimated as
\begin{equation}
 U_{\text{contact}}(\bm{r}) 
  =   \epsilon \left(1 - \frac{a_{0}}{\sqrt{2} \sigma_{0}} \right)^{5/2}
  \approx \epsilon \left[1 - \frac{(2 \pi / 3)^{1/3}}{\sqrt{2}} \right]^{5/2}.
\end{equation}
By the volume compression, the lattice constant is changed from $a_{0}$ to $a = a_{0} (\rho_{0} / \rho)^{1/3}$.
This changes the distance between neighboring particles by the factor $(\rho_{0} / \rho)^{1/3}$.
The contact potential is changed as
\begin{equation}
 U_{\text{contact}}((\rho_{0} / \rho)^{1/3} \bm{r})
  \approx \epsilon \left[1 - \frac{(2 \pi / 3)^{1/3}}{\sqrt{2}} \left(\frac{\rho_{0}}{\rho}\right)^{1/3} \right]^{5/2}.
\end{equation}
The energy increase per unit volume is estimated to be
\begin{equation}
 \label{compress_energy_meso}
\begin{split}
  u_{\text{compress}} 
 & \approx \frac{12 \epsilon}{a_{0}^{3}}
  \left[
   \left[1 - \frac{(2 \pi / 3)^{1/3}}{\sqrt{2}} \left(\frac{\rho_{0}}{\rho}\right)^{1/3} \right]^{5/2}
   - \left[1 - \frac{(2 \pi / 3)^{1/3}}{\sqrt{2}} \right]^{5/2}
  \right] \\
 & \approx 3 \rho_{0} \epsilon
    \left[1 - \frac{(2 \pi / 3)^{1/3}}{\sqrt{2}} \right]^{1/2}
  \left[
 - 2 \left(\frac{\rho_{0}}{\rho} - 1 \right)
 + \left(\frac{\rho_{0}}{\rho} - 1 \right)^{2}
  \right] 
\end{split}
\end{equation}
If we concentrate on the $\rho$ dependence eq~\eqref{compress_energy_meso},
we find that the energy can be separated into two parts: terms
proportional to $(\rho_{0} / \rho - 1)$ and 
$(\rho_{0} / \rho - 1)^{2}$.
The term which is proportional to $(\rho_{0} / \rho - 1)$ would be
interpreted as the contribution of the pressure. We assume that this
term is balanced to the external pressure, and simply ignore it.
Then the term which is proportional to $(\rho_{0} / \rho  - 1)^{2}$ corresponds
to the compressive elastic energy.
From eqs~\eqref{compress_energy_macro} and \eqref{compress_energy_meso},
we have the following estimate:
\begin{equation}
 \label{contact_potential_parameter_estimate}
 \epsilon
    \approx \frac{1}{6 \rho_{0}} \left[1 - \frac{(2 \pi / 3)^{1/3}}{\sqrt{2}} \right]^{-1/2} K_{a}
\end{equation}

By using eqs \eqref{bbond_potential_parameter_estimate},
\eqref{tilt_bend_potentials_parameter_estimate},
\eqref{dbond_potential_parameter_estimate}, and \eqref{contact_potential_parameter_estimate},
 we can estimate other potential parameters such as
$k_{\text{D-bond}}$ in the dimensionless unit.
By using $K_{a} \approx 1 \text{GPa}$,
$G_{\text{s}} \approx 1 \text{GPa}$, and $h \approx \sigma_{0} \approx 10 \text{nm}$,
we have the following estimate for $k_{\text{B-bond}}$ in the dimensionless unit:
\begin{equation}
 k_{\text{B-bond}}
  \approx 8.2.
\end{equation}
Then we can estimate other potential parameters in the dimensionless units:
\begin{equation}
 k_{\text{bend}} \approx k_{\text{tilt}} \approx 0.63 \sim 0.80, \qquad
  k_{\text{D-bond}} \approx 1.1 \times 10^{-2} z_{0}^{-1}.
\end{equation}
The temperature in the dimensionless can be also estimated via eq~\eqref{contact_potential_parameter_estimate}.
Eq~\eqref{contact_potential_parameter_estimate} gives $\epsilon \approx 2.83 \times 10^{-13}\text{J}$.
For $T = 300\text{K}$, we have the temperature as
\begin{equation}
 T_{\text{eff}} \approx \frac{1.38 \times 10^{-23} \text{J}/\text{K} \times 300 \text{K}}{2.83 \times 10^{-16} \text{J}}
  \approx 1.5 \times 10^{-5}.
\end{equation}
Thus the temperature of the system is expected to be very low.
We can estimate the dimensionless temperature in another way.
The potential parameter for ductile bonds can be related to the temperature
as $k_{\text{D-bond}} = 3 k_{B} T / \bar{Q}^{2}$. Then, by setting $\bar{Q} \approx \sigma_{0}$,
we have
\begin{equation}
 T_{\text{eff}} \approx \frac{1}{3} k_{\text{D-bond}} \approx 3.7 \times 10^{-3} z_{0}^{-1}.
\end{equation}
Again, the temperature is expected to be very low.

We should note that these estimates are very rough and not accurate.
Nevertheless, they will be still useful for simulations.
The potential parameters used in the main text are chosen based on
these estimates, and they reproduce various characteristic behaviors
of crystalline polymer solids well.

\section{Average over Lamellar Stacking Directions}
\label{average_over_lamellar_stacking_directions}

In this appendix, we show the method to calculate the average
over different lamellar stacking directions.
A macroscopic specimen will be divided into small mesoscopic regions
so that the lamellar stacking direction in one mesoscopic region
is almost constant. As a rough approximation, we assume that
all the mesoscopic regions are deformed affinely when the specimen
is deformed. Then macroscopic physical quantities can be interpreted
as the averages over different lamellar stacking directions.

We express the unit normal vector to the lamellar stacking direction
of a domain as $\bm{n}$ ($|\bm{n}| = 1$). If the lamellar stacking direction is random,
$\bm{n}$ can be interpreted as a random variable of which probability
distribution is given as $P(\bm{n}) = 1 / 4 \pi$.
Because we are considering the uniaxial elongation, it is convenient to
use the spherical coordinates:
\begin{equation}
 n_{x} = \cos \varphi \sin \vartheta, \qquad n_{y} = \sin \varphi \sin \vartheta,
  \qquad n_{z} = \cos \vartheta,
\end{equation}
where $0 < \varphi \le 2 \pi$ and $0 \le \vartheta \le \phi$.
If a physical quantity $A(\vartheta)$ depends only on $\vartheta$, and 
has the symmetry $A(\pi - \vartheta) = A(\vartheta)$, the 
average over $\bm{u}$ is calculated as
\begin{equation}
 \label{physical_quantity_average_over_theta}
  \bar{A}  = \int_{0}^{\pi} d\vartheta \int_{0}^{2 \pi} d\varphi \, \sin \vartheta \frac{1}{4 \pi} A(\vartheta) \\
   = \frac{1}{2} \int_{0}^{\pi/2} d\vartheta  \, \sin \vartheta A(\vartheta).
\end{equation}
From the symmetry, we have limited the range for $\vartheta$ to
$0 \le \vartheta \le \pi / 2$.

In simulations, we can calculate values only for limited number of $\vartheta$.
In this work, we use $\vartheta = \theta_{i} = i \pi / 12$ ($i = 0,1,2,\dots, 6$).
We express the physical quantity calculated for the $i$-th direction as
$A_{i}$.
Then we approximate eq~\eqref{physical_quantity_average_over_theta} as
\begin{equation}
 \label{physical_quantity_average_over_theta_approx}
  \bar{A} \approx  \sum_{i = 0}^{6} \int_{(\theta_{i} + \theta_{i - 1}) / 2}^{(\theta_{i + 1} + \theta_{i}) / 2} d\vartheta \, \sin \theta_{i} A_{i},
\end{equation}
with $\theta_{-1} = \theta_{0}$ and $\theta_{7} = \theta_{6}$.
Due to the factor $\sin \theta_{i}$ in the integrand,
the weight depends on the direction rather strongly. We can define the
weight factor for the $i$-th direction as
\begin{equation}
 \label{weight_factor_for_average}
 W_{i} = \int_{(\theta_{i} + \theta_{i - 1}) / 2}^{(\theta_{i + 1} + \theta_{i}) / 2} d\vartheta \, \sin \theta_{i}
  = \cos\left(\frac{\theta_{i} + \theta_{i - 1}}{2}\right) - \cos\left(\frac{\theta_{i + 1} - \theta_{i}}{2}\right).
\end{equation}
Then eq~\eqref{physical_quantity_average_over_theta_approx} can be simply
rewritten as
\begin{equation}
 \label{physical_quantity_average_over_theta_approx_simplified}
  \bar{A} \approx  \sum_{i = 0}^{6} W_{i} A_{i}.
\end{equation}
Numerical calculations give
$W_{0} = 0.0085551$,
$W_{1} = 0.0675653$,
$W_{2} = 0.1305262$,
$W_{3} = 0.1845919$,
$W_{4} = 0.2260780$,
$W_{5} = 0.2521572$, and
$W_{6} = 0.1305262$.
As we explained, the weight factor depends strongly on the direction.
The contributions of $\theta_{0}$ and $\theta_{1}$ to the average are small.

\section*{Supporting Information}
Additional elongation simulation data;
high-resolution snapshots, simulation movies,
two-dimensional power spectra of non-affine displacement, 
broken B-bond fractions, stress-strain curves,
and two-dimensional small-angle scattering patterns.


\begin{mcitethebibliography}{66}
\providecommand*\natexlab[1]{#1}
\providecommand*\mciteSetBstSublistMode[1]{}
\providecommand*\mciteSetBstMaxWidthForm[2]{}
\providecommand*\mciteBstWouldAddEndPuncttrue
  {\def\EndOfBibitem{\unskip.}}
\providecommand*\mciteBstWouldAddEndPunctfalse
  {\let\EndOfBibitem\relax}
\providecommand*\mciteSetBstMidEndSepPunct[3]{}
\providecommand*\mciteSetBstSublistLabelBeginEnd[3]{}
\providecommand*\EndOfBibitem{}
\mciteSetBstSublistMode{f}
\mciteSetBstMaxWidthForm{subitem}{(\alph{mcitesubitemcount})}
\mciteSetBstSublistLabelBeginEnd
  {\mcitemaxwidthsubitemform\space}
  {\relax}
  {\relax}

\bibitem[Strobl(1997)]{Strobl-book}
Strobl,~G. \emph{The Physics of Polymers}, 2nd ed.; Springer: Berlin,
  1997\relax
\mciteBstWouldAddEndPuncttrue
\mciteSetBstMidEndSepPunct{\mcitedefaultmidpunct}
{\mcitedefaultendpunct}{\mcitedefaultseppunct}\relax
\EndOfBibitem
\bibitem[Young and Lovell(2011)Young, and Lovell]{Young-Lovell-book}
Young,~R.~J.; Lovell,~P.~A. \emph{Introduction to Polymers}, 3rd ed.; Taylor
  and Francis: Boca Raton, 2011\relax
\mciteBstWouldAddEndPuncttrue
\mciteSetBstMidEndSepPunct{\mcitedefaultmidpunct}
{\mcitedefaultendpunct}{\mcitedefaultseppunct}\relax
\EndOfBibitem
\bibitem[Lin and Argon(1994)Lin, and Argon]{Lin-Argon-1994}
Lin,~L.; Argon,~A.~S. Structure and plastic deformation of polyethylene.
  \emph{J. Mater. Sci.} \textbf{1994}, \emph{29}, 294--323\relax
\mciteBstWouldAddEndPuncttrue
\mciteSetBstMidEndSepPunct{\mcitedefaultmidpunct}
{\mcitedefaultendpunct}{\mcitedefaultseppunct}\relax
\EndOfBibitem
\bibitem[Bowden and Young(1974)Bowden, and Young]{Bowden-Young-1974}
Bowden,~P.~B.; Young,~R.~J. Deformation mechanisms in crystalline polymers.
  \emph{J. Mater. Sci.} \textbf{1974}, \emph{9}, 2034--2051\relax
\mciteBstWouldAddEndPuncttrue
\mciteSetBstMidEndSepPunct{\mcitedefaultmidpunct}
{\mcitedefaultendpunct}{\mcitedefaultseppunct}\relax
\EndOfBibitem
\bibitem[Peterlin(1987)]{Peterlin-1987}
Peterlin,~A. Drawing and extrusion of semi-crystalline polymers. \emph{Collid
  Polym. Sci.} \textbf{1987}, \emph{265}, 357--382\relax
\mciteBstWouldAddEndPuncttrue
\mciteSetBstMidEndSepPunct{\mcitedefaultmidpunct}
{\mcitedefaultendpunct}{\mcitedefaultseppunct}\relax
\EndOfBibitem
\bibitem[Nitta and Takayanagi(2000)Nitta, and
  Takayanagi]{Nitta-Takayanagi-2000}
Nitta,~K.; Takayanagi,~M. Tensile Yield of Isotactic Polypropylene in Terms of
  a Lamellar-Cluster Model. \emph{J. Polym. Sci. B: Polym. Phys.}
  \textbf{2000}, \emph{38}, 1037--1044\relax
\mciteBstWouldAddEndPuncttrue
\mciteSetBstMidEndSepPunct{\mcitedefaultmidpunct}
{\mcitedefaultendpunct}{\mcitedefaultseppunct}\relax
\EndOfBibitem
\bibitem[Nitta and Takayanagi(2003)Nitta, and
  Takayanagi]{Nitta-Takayanagi-2003}
Nitta,~K.; Takayanagi,~M. Novel Proposal of Lamellar Clustering Process for
  Elucidation of Tensile Yield Behavior of Linear Polyethylenes. \emph{J.
  Macromol. Sci. B} \textbf{2003}, \emph{42}, 107--126\relax
\mciteBstWouldAddEndPuncttrue
\mciteSetBstMidEndSepPunct{\mcitedefaultmidpunct}
{\mcitedefaultendpunct}{\mcitedefaultseppunct}\relax
\EndOfBibitem
\bibitem[Tashiro(1993)]{Tashiro-1993}
Tashiro,~K. Molecular theory of mechanical properties of crystalline polymers.
  \emph{Prog. Polym. Sci.} \textbf{1993}, \emph{18}, 377--435\relax
\mciteBstWouldAddEndPuncttrue
\mciteSetBstMidEndSepPunct{\mcitedefaultmidpunct}
{\mcitedefaultendpunct}{\mcitedefaultseppunct}\relax
\EndOfBibitem
\bibitem[Tashiro and Sasaki(2003)Tashiro, and Sasaki]{Tashiro-Sasaki-2003}
Tashiro,~K.; Sasaki,~S. Structural changes in the ordering process of polymers
  as studied by an organized combination of the various measurement techniques.
  \emph{Prog. Polym. Sci.} \textbf{2003}, \emph{28}, 451--519\relax
\mciteBstWouldAddEndPuncttrue
\mciteSetBstMidEndSepPunct{\mcitedefaultmidpunct}
{\mcitedefaultendpunct}{\mcitedefaultseppunct}\relax
\EndOfBibitem
\bibitem[L\'{o}pez-Barr\'{o}n \latin{et~al.}(2017)L\'{o}pez-Barr\'{o}n, Zeng,
  Schaefer, Eberle, Lodge, and
  Bates]{LopezBarron-Zeng-Schaefer-Eberle-Lodge-Bates-2017}
L\'{o}pez-Barr\'{o}n,~C.~R.; Zeng,~Y.; Schaefer,~J.~J.; Eberle,~A. P.~R.;
  Lodge,~T.~P.; Bates,~F.~S. Molecular Alignment in Polyethylene during Cold
  Drawing Using In-Situ SANS and Raman Spectroscopy. \emph{Macromolecules}
  \textbf{2017}, \emph{50}, 3627--3636\relax
\mciteBstWouldAddEndPuncttrue
\mciteSetBstMidEndSepPunct{\mcitedefaultmidpunct}
{\mcitedefaultendpunct}{\mcitedefaultseppunct}\relax
\EndOfBibitem
\bibitem[Butler and Donald(1998)Butler, and Donald]{Butler-Donald-1998}
Butler,~M.~F.; Donald,~A.~M. A Real-Time Simultaneous Small- and Wide-Angle
  X-ray Scattering Study of in Situ Polyethylene Deformation at Elevated
  Temperatures. \emph{Macromolecules} \textbf{1998}, \emph{31},
  6234--6249\relax
\mciteBstWouldAddEndPuncttrue
\mciteSetBstMidEndSepPunct{\mcitedefaultmidpunct}
{\mcitedefaultendpunct}{\mcitedefaultseppunct}\relax
\EndOfBibitem
\bibitem[Jiang \latin{et~al.}(2007)Jiang, Tang, Men, Enderle, Lilge, Roth,
  Gehrke, and Rieger]{Jiang-Tang-Men-Enderle-Lilge-Roth-Gehrke-Rieger-2007}
Jiang,~Z.; Tang,~Y.; Men,~Y.; Enderle,~H.-F.; Lilge,~D.; Roth,~S.~V.;
  Gehrke,~R.; Rieger,~J. Structural Evolution of Tensile-Deformed High-Density
  Polyethylene during Annealing: Scanning Synchrotron Small-Angle X-ray
  Scattering Study. \emph{Macromolecules} \textbf{2007}, \emph{40},
  7263--7269\relax
\mciteBstWouldAddEndPuncttrue
\mciteSetBstMidEndSepPunct{\mcitedefaultmidpunct}
{\mcitedefaultendpunct}{\mcitedefaultseppunct}\relax
\EndOfBibitem
\bibitem[Millot \latin{et~al.}(2017)Millot, S\'{e}gu\'{e}la, Lame, Fillot,
  Rochas, and Sotta]{Millot-Seguela-Lame-Fillot-Rochas-Sotta-2017}
Millot,~C.; S\'{e}gu\'{e}la,~R.; Lame,~O.; Fillot,~L.-A.; Rochas,~C.; Sotta,~P.
  Tensile Deformation of Bulk Polyamide 6 in the Preyield Strain Range.
  Micro-Macro Strain Relationships via in Situ SAXS and WAXS.
  \emph{Macromolecules} \textbf{2017}, \emph{50}, 1541--1553\relax
\mciteBstWouldAddEndPuncttrue
\mciteSetBstMidEndSepPunct{\mcitedefaultmidpunct}
{\mcitedefaultendpunct}{\mcitedefaultseppunct}\relax
\EndOfBibitem
\bibitem[Kishimoto \latin{et~al.}(2020)Kishimoto, Mita, Ogawa, and
  Takenaka]{Kishimoto-Mita-Ogawa-Takenaka-2020}
Kishimoto,~M.; Mita,~K.; Ogawa,~H.; Takenaka,~M. Effect of Submicron Structures
  on the Mechanical Behavior of Polyethylene. \emph{Macromolecules}
  \textbf{2020}, \emph{53}, 9097--9107\relax
\mciteBstWouldAddEndPuncttrue
\mciteSetBstMidEndSepPunct{\mcitedefaultmidpunct}
{\mcitedefaultendpunct}{\mcitedefaultseppunct}\relax
\EndOfBibitem
\bibitem[Siesler(1980)]{Siesler-1980}
Siesler,~H.~W. Fourier transform infrared (FTIR) spectroscopy in polymer
  research. \emph{J. Macromol. Struct.} \textbf{1980}, \emph{59}, 15--37\relax
\mciteBstWouldAddEndPuncttrue
\mciteSetBstMidEndSepPunct{\mcitedefaultmidpunct}
{\mcitedefaultendpunct}{\mcitedefaultseppunct}\relax
\EndOfBibitem
\bibitem[Song \latin{et~al.}(2003)Song, Nitta, and
  Nemoto]{Song-Nitta-Nemoto-2003}
Song,~Y.; Nitta,~K.; Nemoto,~N. Molecular Orientations and True Stress-Strain
  Relationship in Isotactic Polypropylene Film. \emph{Macromolecules}
  \textbf{2003}, \emph{36}, 8066--8073\relax
\mciteBstWouldAddEndPuncttrue
\mciteSetBstMidEndSepPunct{\mcitedefaultmidpunct}
{\mcitedefaultendpunct}{\mcitedefaultseppunct}\relax
\EndOfBibitem
\bibitem[Kida \latin{et~al.}(2015)Kida, Oku, Hiejima, and
  Nitta]{Kida-Oku-Hiejima-Nitta-2015}
Kida,~T.; Oku,~T.; Hiejima,~Y.; Nitta,~K. Deformation mechanism of high-density
  polyethylene probed by in situ Raman spectroscopy. \emph{Polymer}
  \textbf{2015}, \emph{58}, 88--95\relax
\mciteBstWouldAddEndPuncttrue
\mciteSetBstMidEndSepPunct{\mcitedefaultmidpunct}
{\mcitedefaultendpunct}{\mcitedefaultseppunct}\relax
\EndOfBibitem
\bibitem[Kida \latin{et~al.}(2021)Kida, Hiejima, and
  Nitta]{Kida-Hiejima-Nitta-2021}
Kida,~T.; Hiejima,~Y.; Nitta,~K. Microstructural Interpretation of Influences
  of Molecular Weight on the Tensile Properties of High-Density Polyethylene
  Solids Using Rheo-Raman Spectroscopy. \emph{Macromolecules} \textbf{2021},
  \emph{54}, 225--234\relax
\mciteBstWouldAddEndPuncttrue
\mciteSetBstMidEndSepPunct{\mcitedefaultmidpunct}
{\mcitedefaultendpunct}{\mcitedefaultseppunct}\relax
\EndOfBibitem
\bibitem[Kida(2022)]{Kida-2022}
Kida,~T. Raman Spectroscopic Analyses of Structure-Mechanical Properties
  Relationship of Crystalline Polyolefin Materials. \emph{Nihon Reoroji
  Gakkaishi} \textbf{2022}, \emph{50}, 21--29\relax
\mciteBstWouldAddEndPuncttrue
\mciteSetBstMidEndSepPunct{\mcitedefaultmidpunct}
{\mcitedefaultendpunct}{\mcitedefaultseppunct}\relax
\EndOfBibitem
\bibitem[Lee and Rutledge(2011)Lee, and Rutledge]{Lee-Rutledge-2011}
Lee,~S.; Rutledge,~G.~C. Plastic Deformation of Semicrystalline Polyethylene by
  Molecular Simulation. \emph{Macromolecules} \textbf{2011}, \emph{44},
  3096--3108\relax
\mciteBstWouldAddEndPuncttrue
\mciteSetBstMidEndSepPunct{\mcitedefaultmidpunct}
{\mcitedefaultendpunct}{\mcitedefaultseppunct}\relax
\EndOfBibitem
\bibitem[Kim \latin{et~al.}(2014)Kim, Locker, and
  Rutledge]{Kim-Locker-Rutledge-2014}
Kim,~J.~M.; Locker,~R.; Rutledge,~G.~C. Plastic Deformation of Semicrystalline
  Polyethylene under Extension, Compression, and Shear Using Molecular Dynamics
  Simulation. \emph{Macromolecules} \textbf{2014}, \emph{47}, 2515--2528\relax
\mciteBstWouldAddEndPuncttrue
\mciteSetBstMidEndSepPunct{\mcitedefaultmidpunct}
{\mcitedefaultendpunct}{\mcitedefaultseppunct}\relax
\EndOfBibitem
\bibitem[Jabbari-Farouji \latin{et~al.}(2015)Jabbari-Farouji, Rottler, Lame,
  Makke, Perez, and
  Barrat]{JabbariFarouji-Rottler-Lame-Makke-Perez-Barrat-2015}
Jabbari-Farouji,~S.; Rottler,~J.; Lame,~O.; Makke,~A.; Perez,~M.; Barrat,~J.-L.
  Plastic Deformation Mechanisms of Semicrystalline and Amorphous Polymers.
  \emph{ACS Macro Letters} \textbf{2015}, \emph{4}, 147--150\relax
\mciteBstWouldAddEndPuncttrue
\mciteSetBstMidEndSepPunct{\mcitedefaultmidpunct}
{\mcitedefaultendpunct}{\mcitedefaultseppunct}\relax
\EndOfBibitem
\bibitem[Yeh \latin{et~al.}(2017)Yeh, Lenhart, Rutledge, and
  Andzelm]{Yeh-Lenhart-Rutledge-Andzelm-2017}
Yeh,~I.-C.; Lenhart,~J.~L.; Rutledge,~G.~C.; Andzelm,~J.~W. Molecular Dynamics
  Simulation of the Effects of Layer Thickness and Chain Tilt on Tensile
  Deformation Mechanisms of Semicrystalline Polyethylene. \emph{Macromolecules}
  \textbf{2017}, \emph{50}, 1700--1712\relax
\mciteBstWouldAddEndPuncttrue
\mciteSetBstMidEndSepPunct{\mcitedefaultmidpunct}
{\mcitedefaultendpunct}{\mcitedefaultseppunct}\relax
\EndOfBibitem
\bibitem[Higuchi and Kubo(2017)Higuchi, and Kubo]{Higuchi-Kubo-2017}
Higuchi,~Y.; Kubo,~M. Deformation and fracture processes of a lamellar
  structure in polyethylene at the molecular level by a coarse-grained
  molecular dynamics simulation. \emph{Macromolecules} \textbf{2017},
  \emph{50}, 3690--3702\relax
\mciteBstWouldAddEndPuncttrue
\mciteSetBstMidEndSepPunct{\mcitedefaultmidpunct}
{\mcitedefaultendpunct}{\mcitedefaultseppunct}\relax
\EndOfBibitem
\bibitem[Olsson \latin{et~al.}(2018)Olsson, in't Veld, Andreasson, Bergvall,
  Jutemar, Petersson, Rutledge, and
  Kroon]{Olsson-intVeld-Andreasson-Bergvall-Jutemar-Petersson-Rutledge-Kroon-2018}
Olsson,~P.~A.; in't Veld,~P.~J.; Andreasson,~E.; Bergvall,~E.; Jutemar,~E.~P.;
  Petersson,~V.; Rutledge,~G.~C.; Kroon,~M. All-atomic and coarse-grained
  molecular dynamics investigation of deformation in semi-crystalline lamellar
  polyethylene. \emph{Polymer} \textbf{2018}, \emph{153}, 305--316\relax
\mciteBstWouldAddEndPuncttrue
\mciteSetBstMidEndSepPunct{\mcitedefaultmidpunct}
{\mcitedefaultendpunct}{\mcitedefaultseppunct}\relax
\EndOfBibitem
\bibitem[Ranganathan \latin{et~al.}(2020)Ranganathan, Kumar, Brayton,
  Kr\"{o}ger, and Rutledge]{Ranganathan-Kumar-Brayton-Kroger-Rutledge-2020}
Ranganathan,~R.; Kumar,~V.; Brayton,~A.~L.; Kr\"{o}ger,~M.; Rutledge,~G.~C.
  Atomistic Modeling of Plastic Deformation in Semicrystalline Polyethylene:
  Role of Interphase Topology, Entanglements, and Chain Dynamics.
  \emph{Macromolecules} \textbf{2020}, \emph{53}, 4605--4617\relax
\mciteBstWouldAddEndPuncttrue
\mciteSetBstMidEndSepPunct{\mcitedefaultmidpunct}
{\mcitedefaultendpunct}{\mcitedefaultseppunct}\relax
\EndOfBibitem
\bibitem[Hagita \latin{et~al.}(2024)Hagita, Yamamoto, Saito, and
  Abe]{Hagita-Yamamoto-Saito-Abe-2024}
Hagita,~K.; Yamamoto,~T.; Saito,~H.; Abe,~E. Chain-Level Analysis of Reinforced
  Polyethylene through Stretch-Induced Crystallization. \emph{ACS Macro
  Letters} \textbf{2024}, \emph{13}, 247--251\relax
\mciteBstWouldAddEndPuncttrue
\mciteSetBstMidEndSepPunct{\mcitedefaultmidpunct}
{\mcitedefaultendpunct}{\mcitedefaultseppunct}\relax
\EndOfBibitem
\bibitem[Uneyama(2020)]{Uneyama-2020}
Uneyama,~T. Coarse-graining of microscopic dynamics into a mesoscopic transient
  potential model. \emph{Phys. Rev. E} \textbf{2020}, \emph{101}, 032106\relax
\mciteBstWouldAddEndPuncttrue
\mciteSetBstMidEndSepPunct{\mcitedefaultmidpunct}
{\mcitedefaultendpunct}{\mcitedefaultseppunct}\relax
\EndOfBibitem
\bibitem[Uneyama(2022)]{Uneyama-2022}
Uneyama,~T. Application of projection operator method to coarse-grained
  dynamics with transient potential. \emph{Phys. Rev. E} \textbf{2022},
  \emph{105}, 044117\relax
\mciteBstWouldAddEndPuncttrue
\mciteSetBstMidEndSepPunct{\mcitedefaultmidpunct}
{\mcitedefaultendpunct}{\mcitedefaultseppunct}\relax
\EndOfBibitem
\bibitem[Strobl(2000)]{Strolb-2000}
Strobl,~G. From the melt via mesomorphic and granular crystalline layers to
  lamellar crystallites: A major route followed in polymer crystallization?
  \emph{Eru. Phys. J. E} \textbf{2000}, \emph{3}, 163--183\relax
\mciteBstWouldAddEndPuncttrue
\mciteSetBstMidEndSepPunct{\mcitedefaultmidpunct}
{\mcitedefaultendpunct}{\mcitedefaultseppunct}\relax
\EndOfBibitem
\bibitem[Uneyama \latin{et~al.}(2014)Uneyama, Miyata, and
  Nitta]{Uneyama-Miyata-Nitta-2014}
Uneyama,~T.; Miyata,~T.; Nitta,~K. Self-consistent field model simulations for
  statistics of amorphous polymer chains in crystalline lamellar structures.
  \emph{J. Chem. Phys.} \textbf{2014}, \emph{141}, 164906\relax
\mciteBstWouldAddEndPuncttrue
\mciteSetBstMidEndSepPunct{\mcitedefaultmidpunct}
{\mcitedefaultendpunct}{\mcitedefaultseppunct}\relax
\EndOfBibitem
\bibitem[van Duin \latin{et~al.}(2001)van Duin, Dasgupta, Lorant, and
  III]{vanDuin-Dasgupta-Lorant-Goddard-2001}
van Duin,~A. C.~T.; Dasgupta,~S.; Lorant,~F.; III,~W. A.~G. ReaxFF: A Reactive
  Force Field for Hydrocarbons. \emph{J. Phys. Chem. A} \textbf{2001},
  \emph{105}, 9369--9409\relax
\mciteBstWouldAddEndPuncttrue
\mciteSetBstMidEndSepPunct{\mcitedefaultmidpunct}
{\mcitedefaultendpunct}{\mcitedefaultseppunct}\relax
\EndOfBibitem
\bibitem[Landau and Lifshitz(1986)Landau, and
  Lifshitz]{Landau-Lifshitz-elasticity-book}
Landau,~L.~D.; Lifshitz,~E.~M. \emph{Theory of Elasticity}, 3rd ed.;
  Butterworth-Heinemann: Oxford, 1986\relax
\mciteBstWouldAddEndPuncttrue
\mciteSetBstMidEndSepPunct{\mcitedefaultmidpunct}
{\mcitedefaultendpunct}{\mcitedefaultseppunct}\relax
\EndOfBibitem
\bibitem[P\`{a}mies \latin{et~al.}(2009)P\`{a}mies, Cacciuto, and
  Frenkel]{Pamies-Cacciuto-Frenkel-2009}
P\`{a}mies,~J.~C.; Cacciuto,~A.; Frenkel,~D. Phase diagram of Hertzian spheres.
  \emph{J. Chem. Phys.} \textbf{2009}, \emph{131}, 044514\relax
\mciteBstWouldAddEndPuncttrue
\mciteSetBstMidEndSepPunct{\mcitedefaultmidpunct}
{\mcitedefaultendpunct}{\mcitedefaultseppunct}\relax
\EndOfBibitem
\bibitem[Doi and Edwards(1986)Doi, and Edwards]{Doi-Edwards-book}
Doi,~M.; Edwards,~S.~F. \emph{The Theory of Polymer Dynamics}; Oxford
  University Press: Oxford, 1986\relax
\mciteBstWouldAddEndPuncttrue
\mciteSetBstMidEndSepPunct{\mcitedefaultmidpunct}
{\mcitedefaultendpunct}{\mcitedefaultseppunct}\relax
\EndOfBibitem
\bibitem[Noguchi(2011)]{Noguchi-2011}
Noguchi,~H. Solvent-free coarse-grained lipid model for large-scale
  simulations. \emph{J. Chem. Phys.} \textbf{2011}, \emph{134}, 055101\relax
\mciteBstWouldAddEndPuncttrue
\mciteSetBstMidEndSepPunct{\mcitedefaultmidpunct}
{\mcitedefaultendpunct}{\mcitedefaultseppunct}\relax
\EndOfBibitem
\bibitem[Kuzkin and Asonov(2012)Kuzkin, and Asonov]{Kuzkin-Asonov-2012}
Kuzkin,~V.~A.; Asonov,~I.~E. Vector-based model of elastic bonds for simulation
  of granular solids. \emph{Phys. Rev. E} \textbf{2012}, \emph{86},
  051301\relax
\mciteBstWouldAddEndPuncttrue
\mciteSetBstMidEndSepPunct{\mcitedefaultmidpunct}
{\mcitedefaultendpunct}{\mcitedefaultseppunct}\relax
\EndOfBibitem
\bibitem[Kuzkin and Krivtsov(2017)Kuzkin, and Krivtsov]{Kuzkin-Krivtsov-2017}
Kuzkin,~V.~A.; Krivtsov,~A.~M. Enhanced vector-based model for elastic bonds in
  solids. \emph{Lett. Mater.} \textbf{2017}, \emph{7}, 455--458\relax
\mciteBstWouldAddEndPuncttrue
\mciteSetBstMidEndSepPunct{\mcitedefaultmidpunct}
{\mcitedefaultendpunct}{\mcitedefaultseppunct}\relax
\EndOfBibitem
\bibitem[Enomoto \latin{et~al.}(2024)Enomoto, Ishida, Doi, Uneyama, and
  Masubuchi]{Enomoto-Ishida-Doi-Uneyama-Masubuchi-2024}
Enomoto,~K.; Ishida,~T.; Doi,~Y.; Uneyama,~T.; Masubuchi,~Y. Extension of
  moving particle simulation by introducing rotational degrees of freedom for
  dilute fiber suspensions. \emph{Int. J. Numer. Meth. Fluids} \textbf{2024},
  \emph{96}, 125--137\relax
\mciteBstWouldAddEndPuncttrue
\mciteSetBstMidEndSepPunct{\mcitedefaultmidpunct}
{\mcitedefaultendpunct}{\mcitedefaultseppunct}\relax
\EndOfBibitem
\bibitem[Tao \latin{et~al.}(2006)Tao, den Otter, Dhont, and
  Briels]{Tao-denOtter-Dhont-Briels-2006}
Tao,~Y.-G.; den Otter,~W.~K.; Dhont,~J. K.~G.; Briels,~W.~J. Isotropic-nematic
  spinodals of rigid long thin rodlike colloids by event-driven Brownian
  dynamics simulations. \emph{J. Chem. Phys.} \textbf{2006}, \emph{124},
  134906\relax
\mciteBstWouldAddEndPuncttrue
\mciteSetBstMidEndSepPunct{\mcitedefaultmidpunct}
{\mcitedefaultendpunct}{\mcitedefaultseppunct}\relax
\EndOfBibitem
\bibitem[Landau and Lifshitz(1987)Landau, and
  Lifshitz]{Landau-Lifshitz-fluid-book}
Landau,~L.~D.; Lifshitz,~E.~M. \emph{Fluid Mechanics}, 2nd ed.; Pergamon Press:
  Oxford, 1987\relax
\mciteBstWouldAddEndPuncttrue
\mciteSetBstMidEndSepPunct{\mcitedefaultmidpunct}
{\mcitedefaultendpunct}{\mcitedefaultseppunct}\relax
\EndOfBibitem
\bibitem[Gardiner(2004)]{Gardiner-book}
Gardiner,~C.~W. \emph{Handbook of Stochastic Methods}, 3rd ed.; Springer:
  Berlin, 2004\relax
\mciteBstWouldAddEndPuncttrue
\mciteSetBstMidEndSepPunct{\mcitedefaultmidpunct}
{\mcitedefaultendpunct}{\mcitedefaultseppunct}\relax
\EndOfBibitem
\bibitem[Honeycutt(1992)]{Honeycutt-1992}
Honeycutt,~R.~L. Stochastic Runge-Kutta algorithms. I. White noise. \emph{Phys.
  Rev. A} \textbf{1992}, \emph{45}, 600--603\relax
\mciteBstWouldAddEndPuncttrue
\mciteSetBstMidEndSepPunct{\mcitedefaultmidpunct}
{\mcitedefaultendpunct}{\mcitedefaultseppunct}\relax
\EndOfBibitem
\bibitem[Matsumoto and Nishimura(1998)Matsumoto, and
  Nishimura]{Matsumoto-Nishimura-1998}
Matsumoto,~M.; Nishimura,~T. Mersenne twister: a 623-dimensionally
  equidistributed uniform pseudo-random number generator. \emph{ACM Trans.
  Model. Comp. Simul.} \textbf{1998}, \emph{8}, 3--30,
  http://www.math.sci.hiroshima-u.ac.jp/\~{}m-mat/MT/emt.html\relax
\mciteBstWouldAddEndPuncttrue
\mciteSetBstMidEndSepPunct{\mcitedefaultmidpunct}
{\mcitedefaultendpunct}{\mcitedefaultseppunct}\relax
\EndOfBibitem
\bibitem[Devroye(1986)]{Devroye-book}
Devroye,~L. \emph{Non-Uniform Random Variate Generation}; Springer: New York,
  1986\relax
\mciteBstWouldAddEndPuncttrue
\mciteSetBstMidEndSepPunct{\mcitedefaultmidpunct}
{\mcitedefaultendpunct}{\mcitedefaultseppunct}\relax
\EndOfBibitem
\bibitem[Raabe(2004)]{Raabe-2004}
Raabe,~D. Mesoscale simulation of spherulite growth during polymer
  crystallization by use of a cellular automaton. \emph{Acta Mater.}
  \textbf{2004}, \emph{52}, 2653--2664\relax
\mciteBstWouldAddEndPuncttrue
\mciteSetBstMidEndSepPunct{\mcitedefaultmidpunct}
{\mcitedefaultendpunct}{\mcitedefaultseppunct}\relax
\EndOfBibitem
\bibitem[Makke \latin{et~al.}(2012)Makke, Perez, Lame, and
  Barrat]{Makke-Perez-Lame-Barrat-2012}
Makke,~A.; Perez,~M.; Lame,~O.; Barrat,~J.-L. Nanoscale buckling deformation in
  layered copolymer materials. \emph{Proc. Nat. Acad. Sci. USA} \textbf{2012},
  \emph{109}, 680--685\relax
\mciteBstWouldAddEndPuncttrue
\mciteSetBstMidEndSepPunct{\mcitedefaultmidpunct}
{\mcitedefaultendpunct}{\mcitedefaultseppunct}\relax
\EndOfBibitem
\bibitem[Pope and Keller(1975)Pope, and Keller]{Pope-Keller-1975}
Pope,~D.~P.; Keller,~A. Deformation of Oriented Polyethylene. \emph{J. Polym.
  Sci.: Polym. Phys. Ed.} \textbf{1975}, \emph{13}, 533--566\relax
\mciteBstWouldAddEndPuncttrue
\mciteSetBstMidEndSepPunct{\mcitedefaultmidpunct}
{\mcitedefaultendpunct}{\mcitedefaultseppunct}\relax
\EndOfBibitem
\bibitem[Desari and Misra(2003)Desari, and Misra]{Desari-Misra-2003}
Desari,~A.; Misra,~R. D.~K. On the strain rate sensitivity of high density
  polyethylene and polypropylenes. \emph{Materials Science and Engineering A}
  \textbf{2003}, \emph{358}, 356--371\relax
\mciteBstWouldAddEndPuncttrue
\mciteSetBstMidEndSepPunct{\mcitedefaultmidpunct}
{\mcitedefaultendpunct}{\mcitedefaultseppunct}\relax
\EndOfBibitem
\bibitem[van Hecke(2010)]{vanHecke-2009}
van Hecke,~M. Jamming of soft particles: geometry, mechanics, scaling and
  isostaticity. \emph{J. Phys.: Cond. Matt.} \textbf{2010}, \emph{22},
  033101\relax
\mciteBstWouldAddEndPuncttrue
\mciteSetBstMidEndSepPunct{\mcitedefaultmidpunct}
{\mcitedefaultendpunct}{\mcitedefaultseppunct}\relax
\EndOfBibitem
\bibitem[Behringer and Chakraborty(2019)Behringer, and
  Chakraborty]{Behringer-Chakraborty-2019}
Behringer,~R.~P.; Chakraborty,~B. The physics of jamming for granular
  materials: a review. \emph{Rep. Prog. Phys.} \textbf{2019}, \emph{82},
  012601\relax
\mciteBstWouldAddEndPuncttrue
\mciteSetBstMidEndSepPunct{\mcitedefaultmidpunct}
{\mcitedefaultendpunct}{\mcitedefaultseppunct}\relax
\EndOfBibitem
\bibitem[Ericksen(1975)]{Ericksen-1975}
Ericksen,~J.~L. Equilibrium of bars. \emph{J. Elast.} \textbf{1975}, \emph{5},
  191--201\relax
\mciteBstWouldAddEndPuncttrue
\mciteSetBstMidEndSepPunct{\mcitedefaultmidpunct}
{\mcitedefaultendpunct}{\mcitedefaultseppunct}\relax
\EndOfBibitem
\bibitem[Fielding(2007)]{Fielding-2007}
Fielding,~S.~M. Complex dynamics of shear banded flows. \emph{Soft Matter}
  \textbf{2007}, \emph{3}, 1267--1279\relax
\mciteBstWouldAddEndPuncttrue
\mciteSetBstMidEndSepPunct{\mcitedefaultmidpunct}
{\mcitedefaultendpunct}{\mcitedefaultseppunct}\relax
\EndOfBibitem
\bibitem[Sato \latin{et~al.}(2010)Sato, Yuan, and
  Kawakatsu]{Sato-Yuan-Kawakatsu-2010}
Sato,~K.; Yuan,~X.-F.; Kawakatsu,~T. Why does shear banding behave like
  first-order phase transitions? Derivation of a potential from a mechanical
  constitutive model. \emph{Eur. Phys. J. E} \textbf{2010}, \emph{31},
  135--144\relax
\mciteBstWouldAddEndPuncttrue
\mciteSetBstMidEndSepPunct{\mcitedefaultmidpunct}
{\mcitedefaultendpunct}{\mcitedefaultseppunct}\relax
\EndOfBibitem
\bibitem[Landau and Lifshitz(1980)Landau, and
  Lifshitz]{Landau-Lifshitz-statistical-book}
Landau,~L.~D.; Lifshitz,~E.~M. \emph{Statistical Physics Part 1}, 3rd ed.;
  Butterworth-Heinemann: Oxford, 1980\relax
\mciteBstWouldAddEndPuncttrue
\mciteSetBstMidEndSepPunct{\mcitedefaultmidpunct}
{\mcitedefaultendpunct}{\mcitedefaultseppunct}\relax
\EndOfBibitem
\bibitem[Adhikari and Michler(2004)Adhikari, and
  Michler]{Adhikari-Michler-2004}
Adhikari,~R.; Michler,~G.~H. Influence of molecular architecture on morphology
  and micromechanical behavior of styrene/butadiene. \emph{Prog. Polym. Sci.}
  \textbf{2004}, \emph{29}, 949--986\relax
\mciteBstWouldAddEndPuncttrue
\mciteSetBstMidEndSepPunct{\mcitedefaultmidpunct}
{\mcitedefaultendpunct}{\mcitedefaultseppunct}\relax
\EndOfBibitem
\bibitem[Fredrickson(2006)]{Fredrickson-book}
Fredrickson,~G.~H. \emph{The Equilibrium Theory of Inhomogeneous Polymers};
  Oxford University Press: Oxford, 2006\relax
\mciteBstWouldAddEndPuncttrue
\mciteSetBstMidEndSepPunct{\mcitedefaultmidpunct}
{\mcitedefaultendpunct}{\mcitedefaultseppunct}\relax
\EndOfBibitem
\bibitem[Leibler(1980)]{Leibler-1980}
Leibler,~L. Theory of Microphase Separation in Block Copolymers.
  \emph{Macromolecules} \textbf{1980}, \emph{13}, 1602--1617\relax
\mciteBstWouldAddEndPuncttrue
\mciteSetBstMidEndSepPunct{\mcitedefaultmidpunct}
{\mcitedefaultendpunct}{\mcitedefaultseppunct}\relax
\EndOfBibitem
\bibitem[Ohta and Kawasaki(1986)Ohta, and Kawasaki]{Ohta-Kawasaki-1986}
Ohta,~T.; Kawasaki,~K. Equilibrium Morphology of Block Copolymer Melts.
  \emph{Macromolecules} \textbf{1986}, \emph{19}, 2621--2632\relax
\mciteBstWouldAddEndPuncttrue
\mciteSetBstMidEndSepPunct{\mcitedefaultmidpunct}
{\mcitedefaultendpunct}{\mcitedefaultseppunct}\relax
\EndOfBibitem
\bibitem[Matsen and Bates(1996)Matsen, and Bates]{Matsen-Bates-1996}
Matsen,~M.~W.; Bates,~F.~S. Unifying Weak- and Strong-Segregation Block
  Copolymer Theories. \emph{Macromolecules} \textbf{1996}, \emph{29},
  1091--1098\relax
\mciteBstWouldAddEndPuncttrue
\mciteSetBstMidEndSepPunct{\mcitedefaultmidpunct}
{\mcitedefaultendpunct}{\mcitedefaultseppunct}\relax
\EndOfBibitem
\bibitem[M\"uller and Schmid(2005)M\"uller, and Schmid]{Muller-Schmid-2005}
M\"uller,~M.; Schmid,~F. Incorporating Fluctuations and Dynamics in
  Self-Consistent Field Theories for Polymer Blends. \emph{Adv. Polym. Sci.}
  \textbf{2005}, \emph{185}, 1--58\relax
\mciteBstWouldAddEndPuncttrue
\mciteSetBstMidEndSepPunct{\mcitedefaultmidpunct}
{\mcitedefaultendpunct}{\mcitedefaultseppunct}\relax
\EndOfBibitem
\bibitem[Uneyama and Doi(2005)Uneyama, and Doi]{Uneyama-Doi-2005}
Uneyama,~T.; Doi,~M. Density Functional Theory for Block Copolymer Melts and
  Blends. \emph{Macromolecules} \textbf{2005}, \emph{38}, 196--205\relax
\mciteBstWouldAddEndPuncttrue
\mciteSetBstMidEndSepPunct{\mcitedefaultmidpunct}
{\mcitedefaultendpunct}{\mcitedefaultseppunct}\relax
\EndOfBibitem
\bibitem[Aoyagi \latin{et~al.}(2002)Aoyagi, Honda, and
  Doi]{Aoyagi-Honda-Doi-2002}
Aoyagi,~T.; Honda,~T.; Doi,~M. Microstructural study of mechanical properties
  of the ABA triblock copolymer using self-consistent field and molecular
  dynamics. \emph{J. Chem. Phys.} \textbf{2002}, \emph{117}, 8153--8161\relax
\mciteBstWouldAddEndPuncttrue
\mciteSetBstMidEndSepPunct{\mcitedefaultmidpunct}
{\mcitedefaultendpunct}{\mcitedefaultseppunct}\relax
\EndOfBibitem
\bibitem[Makke \latin{et~al.}(2012)Makke, Lame, Perez, and
  Barrat]{Makke-Lame-Perez-Barrat-2012}
Makke,~A.; Lame,~O.; Perez,~M.; Barrat,~J.-L. Influence of Tie and Loop
  Molecules on the Mechanical Properties of Lamellar Block Copolymers.
  \emph{Macromolecules} \textbf{2012}, \emph{45}, 8445--8452\relax
\mciteBstWouldAddEndPuncttrue
\mciteSetBstMidEndSepPunct{\mcitedefaultmidpunct}
{\mcitedefaultendpunct}{\mcitedefaultseppunct}\relax
\EndOfBibitem
\bibitem[Makke \latin{et~al.}(2013)Makke, Lame, Perez, and
  Barrat]{Makke-Lame-Perez-Barrat-2013}
Makke,~A.; Lame,~O.; Perez,~M.; Barrat,~J.-L. Nanoscale Buckling in Lamellar
  Block Copolymers: A Molecular Dynamics Simulation Approach.
  \emph{Macromolecules} \textbf{2013}, \emph{46}, 7853--7864\relax
\mciteBstWouldAddEndPuncttrue
\mciteSetBstMidEndSepPunct{\mcitedefaultmidpunct}
{\mcitedefaultendpunct}{\mcitedefaultseppunct}\relax
\EndOfBibitem
\bibitem[Hagita \latin{et~al.}(2019)Hagita, Akutagawa, Tominaga, and
  Jinnai]{Hagita-Akutagawa-Tominaga-Jinnai-2019}
Hagita,~K.; Akutagawa,~K.; Tominaga,~T.; Jinnai,~H. Scattering patterns and
  stress-strain relations on phase-separated ABA block copolymers under
  uniaxial elongating simulations. \emph{Soft Matter} \textbf{2019}, \emph{15},
  926--936\relax
\mciteBstWouldAddEndPuncttrue
\mciteSetBstMidEndSepPunct{\mcitedefaultmidpunct}
{\mcitedefaultendpunct}{\mcitedefaultseppunct}\relax
\EndOfBibitem
\end{mcitethebibliography}
\providecommand{\latin}[1]{#1}
\makeatletter
\providecommand{\doi}
  {\begingroup\let\do\@makeother\dospecials
  \catcode`\{=1 \catcode`\}=2 \doi@aux}
\providecommand{\doi@aux}[1]{\endgroup\texttt{#1}}
\makeatother
\providecommand*\mcitethebibliography{\thebibliography}
\csname @ifundefined\endcsname{endmcitethebibliography}
  {\let\endmcitethebibliography\endthebibliography}{}

\end{document}